\documentclass[preprint]{aastex}

\usepackage{euscript,epsfig,amsmath,amssymb,amsfonts,latexsym}

\newcommand{\be}[1]{\begin{equation}\label{#1}}
\newcommand{\ee}{\end{equation}}
\newcommand{\ba}[1]{\begin{eqnarray}\label{#1}}
\newcommand{\ea}{\end{eqnarray}}
\newcommand{\rf}[1]{(\ref{#1})}
\newcommand{\nn}{\nonumber}

\shorttitle{Second-order cosmological perturbations} \shortauthors{R. Brilenkov \& M. Eingorn}

\begin{document}

\title{Second-order Cosmological Perturbations Engendered by Point-like Masses}

\author{Ruslan Brilenkov}

\affil{Institute for Astro- and Particle Physics, University of Innsbruck\\ Technikerstrasse 25/8, A-6020 Innsbruck, Austria\\}

\affil{Dipartimento di Fisica e Astronomia `G. Galilei', Universit\`{a} di Padova\\ vicolo dell'Osservatorio 3, 35122 Padova, Italy\\}

\email{ruslan.brilenkov@gmail.com}

\author{Maxim Eingorn}

\affil{North Carolina Central University, CREST and NASA Research Centers\\ Fayetteville st. 1801, Durham, North Carolina 27707, U.S.A.\\}

\email{maxim.eingorn@gmail.com}

\begin{abstract}
In the $\Lambda$CDM framework, presenting nonrelativistic matter inhomogeneities as discrete massive particles, we develop the second-order cosmological
perturbation theory.~Our approach relies on the weak gravitational field limit. The derived equations for the second-order scalar, vector and tensor metric
corrections are suitable at arbitrary distances including regions with nonlinear contrasts of the matter density. We thoroughly verify fulfilment of all Einstein
equations as well as self-consistency of order assignments. In addition, we achieve logical positive results in the Minkowski background limit. Feasible
investigations of the cosmological backreaction manifestations by means of relativistic simulations are also outlined.
\end{abstract}


\

\section{INTRODUCTION}

The conventional Lambda Cold Dark Matter ($\Lambda$CDM) model conforms with the observational data (see, in particular, \cite{Planck15}) and embodies the
mainstream of modern cosmology despite the distressing fact that the nature of dark ingredients of the Universe still remains uncertain. The key assumption, being
typical for this cosmological model as well as its numerous alternatives, is the existence of the homogeneous and isotropic
Friedmann-Lema\^{\i}tre-Robertson-Walker (FLRW) background, which is only slightly perturbed by inhomogeneities inherent in the distribution of the world's
filling material. The following quite natural question arises: can galaxies and their accumulations lead to considerable metric corrections and deeply affect the
average cosmic expansion? The affirmative answer would bring to general recognition of the so-called backreaction (see the reviews by
\cite{Rasanen,Clarkson2,Buchert1,Bolejko} and Refs. therein; also for a recent spirited debate about the magnitude of backreaction effects see, in particular,
\cite{Green,Buchert2} and Refs. therein) and give at least a glimmer of hope to explain the apparent acceleration of expansion without mysterious dark energy.

Even if the answer is negative, it is extremely important to estimate deviations from the FLRW description and confront the theoretical predictions with the
promising outcomes of such future space missions as Euclid \citep{Euclid1,Euclid2}. For this purpose a reliable cosmological perturbation theory should be
developed in the general relativity (GR) framework. Appropriateness at all cosmic scales and non-perturbative treatment of the matter density are those basic
requirements which such a theory must meet. The first-order scheme complying with these reasonable demands and being suitable for relativistic $N$-body
simulations has been successfully constructed for a system of discrete massive particles with nonrelativistic velocities by \cite{Eingorn}. Some predecessors and
their drawbacks are reviewed in \citep{Eingorn} as well.

The current paper is devoted to the generalization of the afore-mentioned first-order approach to the second order with respect to deviations of the metric
coefficients from the corresponding background quantities. This generalization is particularly motivated by the fact that the first order is obviously
insufficient for the comprehensive analysis of the possible backreaction manifestations and the trustworthy prediction of their magnitude. The appeal to the
second order with the purpose of revealing the corresponding observable features is also very promising in the era of precision cosmology (see
\cite{observ1,observ2,observ3} and Refs. therein).

The narration is organized as follows: after reviewing the basic results of \cite{Eingorn} in Section~\ref{sec2}, we switch over to derivation and verification of
equations for the second-order scalar, vector and tensor perturbations in Section~\ref{sec3}. Then, in Section~\ref{sec4}, we focus attention on identification of
the effective average energy density and pressure, and propose the research program aimed at the perturbative computation of the cosmological backreaction
effects. Our achievements are laconically summarized in concluding Section~\ref{sec5}.

\

\section{DISCRETE PICTURE OF THE FIRST-ORDER COSMOLOGICAL PERTURBATIONS AT ALL SCALES}\label{sec2}

\setcounter{equation}{0}

Let us start with an overview of the newly formulated first-order perturbation theory covering all cosmological spatial scales and permitting of nonlinear
contrasts of the matter density \citep{Eingorn}. We confine ourselves to the conventional $\Lambda$CDM model (with zero spatial curvature) and concentrate on
those stages of the Universe evolution when cold matter (dark and baryonic) and dark energy (being represented by the cosmological constant $\Lambda$) dominate
while radiation or relativistic cosmic neutrinos make negligible contributions (see, however, \cite{Ruslan,ekz} for a broad generalization to the multicomponent
case). Then the homogeneous and isotropic cosmological background is described by the unperturbed FLRW metric
\be{1a} ds^2=a^2\left(d\eta^2-\delta_{\alpha\beta}dx^{\alpha}dx^{\beta}\right),\quad \alpha,\beta=1,2,3\, , \ee
where $\eta$ and $x^{\alpha}$, $\alpha=1,2,3$, stand for the conformal time and comoving coordinates, respectively, and the corresponding Friedmann equations for
the scale factor $a(\eta)$:
\be{2a} \frac{3{\mathcal H}^2}{a^2}=\kappa\overline{\varepsilon} + \Lambda,\quad \frac{2{\mathcal H}'+{\mathcal H}^2}{a^2}=\Lambda\, . \ee
Here ${\mathcal H}(\eta)\equiv a'/a$ (with prime denoting the derivative with respect to $\eta$) and $\kappa\equiv 8\pi G_N/c^4$ (with $c$ and $G_N$ representing
the speed of light and Newtonian gravitational constant, respectively). Further, $\varepsilon$ is the energy density of the nonrelativistic pressureless matter,
and the overline indicates averaging.

In the first-order approximation the real inhomogeneous Universe is usually assumed to be described well by the perturbed metric
\citep{Bardeen,Mukhanov,Rubakov,Ruth}
\be{3a} ds^2=a^2\left[\left(1 + 2\Phi\right)d\eta^2 + 2B_{\alpha}dx^{\alpha}d\eta - \left(1- 2\Phi\right)\delta_{\alpha\beta}dx^{\alpha}dx^{\beta}\right]\, . \ee
Here $\Phi(\eta,{\bf r})$ is the scalar perturbation while the spatial vector ${\bf B}(\eta,{\bf r})\equiv\left(B_1,B_2,B_3\right)\equiv\left(B_x,B_y,B_z\right)$
stands for the vector perturbation and satisfies the prevalent gauge condition
\be{4a} \nabla {\bf B}\equiv \delta^{\alpha\beta}\frac{\partial B_{\alpha}}{\partial x^{\beta}}\equiv 0\, .\ee
Similarly to \cite{Clarkson1,Baumann}, we have chosen the popular Poisson gauge but have not yet taken account of the tensor perturbations treating them as
second-order quantities. As pointed out by \cite{Eingorn}, one can in principle allow for the first-order tensor perturbations associated with freely propagating
gravitational waves (with no generator). They have not been explicitly included in \rf{3a} since below we totally neglect their possible contributions to the
sources of second-order metric corrections. At the same time, similarly to \cite{Eingorn} but in contrast to \cite{Clarkson1,Baumann}, the vector perturbation
${\bf B}$ has been already included in \rf{3a} as a first-order quantity since it has a definite nonzero generator (see the right-hand side (rhs) of the equation
\rf{9a} for ${\bf B}$ below).

In \citep{Eingorn} the role of the inhomogeneous gravitational field source belongs to a system of separate nonrelativistic point-like particles with masses
$m_n$, comoving radius-vectors ${\bf r}_n(\eta)$ and comoving peculiar velocities ${\bf\tilde v}_n(\eta)\equiv d{\bf
r}_n/d\eta\equiv(\tilde{v}_n^1,\tilde{v}_n^2,\tilde{v}_n^3)$, and the following expressions for $\Phi$, ${\bf B}$ are derived:
\ba{5a} \Phi&=&\frac{1}{3}-\frac{\kappa c^2}{8\pi a}\sum\limits_n\frac{m_n}{|{\bf r}-{\bf r}_n|}\exp(-q_n)\nn\\
&+&\frac{3\kappa c^2}{8\pi a}{\mathcal H}\sum\limits_n \frac{m_n[{\bf\tilde v}_n({\bf r}-{\bf r}_n)]}{|{\bf r}-{\bf r}_n|}\cdot\frac{1-(1+q_n)\exp(-q_n)}{q_n^2}\,
,\ea
\ba{6a} {\bf B}&=& \frac{\kappa c^2}{8\pi a}\sum\limits_{n}\left[\frac{m_n{\bf\tilde v}_n}{|{\bf r}-{\bf r}_n|}\cdot
\frac{\left(3+2\sqrt{3}q_n+4q_n^2\right)\exp\left(-2q_n/\sqrt{3}\right)-3}{q_n^2}\right.\nn\\
&+&\left.\frac{m_n[{\bf\tilde v}_n({\bf r}-{\bf r}_n)]}{|{\bf r}-{\bf r}_n|^3}({\bf r}-{\bf
r}_n)\cdot\frac{9-\left(9+6\sqrt{3}q_n+4q_n^2\right)\exp\left(-2q_n/\sqrt{3}\right)}{q_n^2}\right]\, .\ea
Here $q_n(\eta,{\bf r}) \equiv a|{\bf r}-{\bf r}_n|/\lambda$, with $\lambda(\eta)\equiv\left[2a^3/\left(3\kappa\overline\rho c^2\right)\right]^{1/2}\sim a^{3/2}$
defining a finite range of Yukawa interaction. Being armed with the observed values of the Hubble constant $H_0\approx68\, \mathrm{km\,s^{-1}\,Mpc^{-1}}$ and the
parameter $\Omega_\mathrm{M}\equiv\kappa\overline\rho c^4/\left(3H_0^2a_0^3\right)\approx0.31$ \citep{Planck15}, where $a_0$ denotes the today's scale factor, one
can easily estimate the current value of the introduced characteristic cutoff scale: $\lambda_0\approx3.7\, \mathrm{Gpc}$ \citep{Eingorn}. An evident relationship
$\overline\varepsilon=\overline\rho c^2/a^3$ establishes linkage between $\overline\varepsilon(\eta)\sim a^{-3}$ and the constant average rest mass density
$\overline\rho$. As regards the corresponding non-averaged quantity, the rest mass density of the analyzed particle system in the comoving coordinates has the
form
\be{7a} \rho(\eta,{\bf r})=\sum\limits_nm_n\delta({\bf r}-{\bf r}_n)=\sum\limits_n\rho_n,\quad \rho_n(\eta,{\bf r})\equiv m_n\delta({\bf r}-{\bf r}_n)\, .\ee

The explicit analytical expressions \rf{5a} and \rf{6a} and their noteworthy features along with the underlying perturbative approach and its numerous physical
implications and advantages are analyzed in detail by \cite{Eingorn}, and it makes no sense to manifest hair-splitting by repeating all deserving results here.
Nevertheless, let us briefly enumerate those facts which are crucial for the clear purpose of the next section.

First of all, we stubbornly adhere to the following well-grounded argumentation (see also \cite{Baumann} for the similar reasoning): the metric corrections and
peculiar particle velocities are assumed to be small at arbitrary distances, however, the smallness of the density contrast $\delta\equiv\delta\rho/\overline\rho$
(where $\delta\rho\equiv\rho-\overline\rho$) is not demanded. For instance, $\rho\gg\overline\rho$ within galaxies, but even for huge density contrasts the metric
is still approximated well by \rf{3a}. Therefore, one can linearize the Einstein equations in the first-order metric perturbations $\Phi$; ${\bf B}$
and velocities ${\bf\tilde v}_n$ without resorting to the unnecessary restrictive inequality $|\delta|\ll1$. Thus, the nonlinear deviation of $\rho$ from its
average value $\overline\rho$, actually occurring at sufficiently small scales, is absolutely unforbidden and fully taken into consideration, as opposed to the
standard first-order relativistic perturbation theory \citep{Mukhanov,Ruth,Rubakov} and its generalization to the second order (see, for instance,
\cite{Bartolo}).

Secondly, the functions \rf{5a} and \rf{6a} are found as exact solutions of the Helmholtz equations (see \cite{Eingorn} for the corresponding Fourier transforms):
\be{8a} \triangle\Phi-\frac{3\kappa\overline\rho c^2}{2a}\Phi = \frac{\kappa c^2}{2a}\delta\rho-\frac{3\kappa c^2}{2a}{\mathcal H}\Xi\, ,\ee
\be{9a} \triangle{\bf B}-\frac{2\kappa\overline\rho c^2}{a}{\bf B}=-\frac{2\kappa c^2}{a}\left(\sum\limits_n\rho_n{\bf\tilde v}_n-\nabla\Xi\right)\, ,\ee
where $\triangle\equiv\delta^{\alpha\beta}\cfrac{\partial^2}{\partial x^{\alpha}\partial x^{\beta}}$ is the Laplace operator in the comoving coordinates, and the
auxiliary function $\Xi(\eta,{\bf r})$ has been introduced:
\be{10a} \Xi=\frac{1}{4\pi}\sum\limits_nm_n\frac{({\bf r}-{\bf r}_n){\bf \tilde v}_n}{|{\bf r}-{\bf r}_n|^3}\, .\ee
Evidently, the presence of the term $\sim\Phi$ in the equation \rf{8a} for the scalar perturbation $\Phi$ is the reason for the Yukawa-type cutoff in the solution
\rf{5a}. The afore-mentioned screening length $\lambda$ is naturally related to the factor in this term: $3\kappa\overline\rho c^2/(2a)\equiv a^2/\lambda^2$. The
contribution $\sim\Phi$ itself as well as the contribution $\sim{\bf B}$ in the equation \rf{9a} arise in view of the fact that the energy-momentum fluctuations,
generating the metric corrections, contain these corrections themselves (see the formulas (2.13) and (2.14) in \citep{Eingorn} as well as the expressions
\rf{22b}--\rf{24b} below).

The vector perturbation ${\bf B}$ \rf{6a} obeys the gauge condition \rf{4a}. In addition, within the adopted accuracy the perturbations \rf{5a} and \rf{6a}
satisfy all remaining linearized Einstein equations, which are reduced to the triplet containing temporal derivatives:
\ba{11a} \Phi'+{\mathcal H}\Phi=-\frac{\kappa c^2}{2a}\Xi,\quad \Phi''+3{\mathcal H}\Phi'+\left(2{\mathcal H}'+{\mathcal H}^2\right)\Phi=0,\quad {\bf
B}'+2{\mathcal H}{\bf B}=0\, . \ea
Finally, the auxiliary function \rf{10a} is found as the exact solution of the Poisson equation:
\be{12a} \triangle\Xi=\nabla\sum\limits_n\rho_n{\bf\tilde v}_n=-\sum\limits_n\rho'_n\, ,\ee
where the continuity equation $\rho'_n+\nabla(\rho_n{\bf\tilde v}_n)=0$, which is satisfied for any $n$-th particle identically, has been employed.

It presents no difficulty to show that the usual theory of hydrodynamical fluctuations emerges in the continuum limit. Indeed, let us momentarily regard
$\rho(\eta,{\bf r})$ as a continuous mass density field and ${\bf\tilde v}(\eta,{\bf r})=\nabla \tilde v^{(\parallel)}+{\bf \tilde v}^{(\perp)}$ as a continuous
velocity field (with $\tilde v^{(\parallel)}(\eta,{\bf r})$ and ${\bf \tilde v}^{(\perp)}(\eta,{\bf r})$ denoting the scalar and vector parts, respectively).
Further, we replace $\Xi$ by $\overline\rho \tilde v^{(\parallel)}$ and $\left(\sum\limits_n\rho_n{\bf\tilde v}_n-\nabla\Xi\right)$ by $\overline{\rho}{\bf \tilde
v}^{(\perp)}$ in the case of linear density fluctuations. Then we make use of the relationship $\varepsilon=\rho c^2/a^3+\left(3\overline\rho c^2/a^3\right)\Phi$
in order to rewrite equations in terms of the energy density field $\varepsilon(\eta,{\bf r})$ (instead of $\rho$). Finally, we introduce covariant spatial
components of the 4-velocity $u_{\beta}=aB_{\beta}-a\tilde v^{\beta}$, $\beta=1,2,3$, and a spatial vector with these components ${\bf u}(\eta,{\bf
r})\equiv\left(u_1,u_2,u_3\right)$ (instead of ${\bf \tilde v}$). We also single out the corresponding scalar and vector parts: ${\bf u}=\nabla
u^{(\parallel)}+{\bf u}^{(\perp)}$, where $u^{(\parallel)}(\eta,{\bf r})=-a\tilde v^{(\parallel)}$ and ${\bf u}^{(\perp)}(\eta,{\bf r})=a{\bf B}-a{\bf \tilde
v}^{(\perp)}$. As a result, the rewritten equations \rf{8a}, \rf{9a}, \rf{11a} become equivalent to those describing hydrodynamical perturbations (see
\cite{Mukhanov}, section 7.3, in the pressureless matter case).

Since the first-order metric corrections inevitably contribute to the second-order ones due to nonlinearity of the Einstein equations, we make the most out of the
enumerated equations for $\Phi$ and ${\bf B}$ as well as $\Xi$ in the very next section.

\

\section{SPLENDORS OF THE SECOND-ORDER THEORY}\label{sec3}

\setcounter{equation}{0}

\subsection{Sources of Perturbations}\label{subsec31}

Switching over to the second-order approximation, we present the metric as
\ba{1b} ds^2&=&a^2\left[\left(1 + 2\Phi\right)d\eta^2 + 2B_{\alpha}dx^{\alpha}d\eta-\left(1- 2\Phi\right)\delta_{\alpha\beta}dx^{\alpha}dx^{\beta}\right]
\nn\\
&+& a^2\left[2\Phi^{(2)}d\eta^2 + 2B_{\alpha}^{(2)}dx^{\alpha}d\eta+\left(2\Psi^{(2)}\delta_{\alpha\beta}+h_{\alpha\beta}\right)dx^{\alpha}dx^{\beta}\right]\, ,
\ea
where $\Phi^{(2)}(\eta,{\bf r})$ and $\Psi^{(2)}(\eta,{\bf r})$ are the second-order scalar perturbations while the spatial vector ${\bf B}^{(2)}(\eta,{\bf
r})\equiv\left(B_1^{(2)},B_2^{(2)},B_3^{(2)}\right)\equiv\left(B_x^{(2)},B_y^{(2)},B_z^{(2)}\right)$ stands for the second-order vector perturbation and satisfies
the same gauge condition as \rf{4a} for ${\bf B}$:
\be{2b} \nabla {\bf B}^{(2)}\equiv\delta^{\alpha\beta}\frac{\partial B^{(2)}_{\alpha}}{\partial x^{\beta}}\equiv0\, .\ee
Similarly to \cite{Clarkson1,Baumann}, we have included the second-order tensor perturbations $h_{\alpha\beta}$, $\alpha,\beta=1,2,3$, in \rf{1b}. They obey the
standard ``transverse-traceless'' gauge conditions:
\be{3b} \delta^{\alpha\beta}\frac{\partial h_{\alpha\gamma}}{\partial x^{\beta}}\equiv0,\quad \delta^{\alpha\beta}h_{\alpha\beta}\equiv0\, .\ee
It should be noted that we decompose the second-order metric corrections into scalar, vector and tensor modes with respect to the unperturbed background. Thus,
for example, the covariant divergence of the total vector ${\bf B}+{\bf B}^{(2)}$, defined via the perturbed spatial metric, does not vanish under the made gauge
choice.

In order to elaborate on the Einstein equations
\be{4b} G_i^k=\kappa T_i^k+\Lambda\delta_i^k,\quad i,k=0,1,2,3\, ,\ee
for the sought-for functions $\Phi^{(2)}$, $\Psi^{(2)}$, ${\bf B}^{(2)}$ and $h_{\alpha\beta}$, first of all it is necessary to calculate the mixed components of
the Einstein tensor $G_i^k$ and matter energy-momentum tensor $T_i^k$ up to the second order in metric corrections and their sources. Let us start with presenting
the left-hand side (lhs) of \rf{4b} as follows (see also \cite{Baumann} for a similar decomposition):
\be{5b} G_i^k=\left(G_i^k\right)^{(0)}+\left(G_i^k\right)^{(1)}+\left(G_i^k\right)^{(2)}+\left(G_i^k\right)^{(\mathrm{II})}\, .\ee
Here $\left(G_i^k\right)^{(0)}$ corresponds to the unperturbed cosmological background metric \rf{1a}. The expressions $\left(G_i^k\right)^{(1)}$ (or, by analogy,
$\left(G_i^k\right)^{(2)}$) are constructed from the terms being linear in the first-order (or second-order) perturbations $\Phi$ and ${\bf B}$ (or $\Phi^{(2)}$,
$\Psi^{(2)}$, ${\bf B}^{(2)}$ and $h_{\alpha\beta}$) and their spatio-temporal derivatives. Finally, $\left(G_i^k\right)^{(\mathrm{II})}$ represent the
second-order quantities containing products of $\Phi$, ${\bf B}$ and their derivatives. We express the temporal derivatives $\Phi'$, $\Phi''$ and ${\bf B}'$ in
$\left(G_i^k\right)^{(\mathrm{II})}$ by means of the functions $\Phi$ and ${\bf B}$ themselves with the help of the triplet \rf{11a}. Thus, for example,
\be{6b} G_{\beta}^0=\left(G_{\beta}^0\right)^{(0)}+\left(G_{\beta}^0\right)^{(1)}+\left(G_{\beta}^0\right)^{(2)}+\left(G_{\beta}^0\right)^{(\mathrm{II})},\quad
\beta=1,2,3\, ,\ee
where
\be{7b} \left(G_{\beta}^0\right)^{(0)}=0\, ,\ee
\be{8b} \left(G_{\beta}^0\right)^{(1)}=\frac{1}{2a^2}\triangle B_{\beta}+ \frac{2}{a^2}\mathcal H\frac{\partial \Phi}{\partial x^{\beta}}
+\frac{2}{a^2}\frac{\partial \Phi'}{\partial x^{\beta}}\, ,\ee
\be{9b} \left(G_{\beta}^0\right)^{(2)}=\frac{1}{2a^2}\triangle B_{\beta}^{(2)}+ \frac{2}{a^2}\mathcal H\frac{\partial \Phi^{(2)}}{\partial x^{\beta}}
+\frac{2}{a^2}\frac{\partial \left(\Psi^{(2)}\right)'}{\partial x^{\beta}}\, ,\ee
\ba{10b} \left(G_{\beta}^0\right)^{(\mathrm{II})}&=&\frac{1}{a^2}B_{\beta}\triangle \Phi +\frac{1}{a^2}\mathcal{H}\frac{\partial \left({\bf
B}^2\right)}{\partial x^{\beta}}-\frac{5}{a^2}\mathcal{H}\frac{\partial \left(\Phi^2\right)}{\partial x^{\beta}}\nn\\
&-&\frac{\kappa c^2}{a^3}\Xi\frac{\partial \Phi}{\partial x^{\beta}}- \frac{1}{a^2}\frac{\partial \Phi}{\partial x^{\alpha}}\frac{\partial B_{\alpha}}{\partial
x^{\beta}}+\frac{1}{a^2}B_{\alpha}\frac{\partial^2 \Phi}{\partial x^{\alpha}\partial x^{\beta}}\, .\ea
Here ${\bf B}^2\equiv B_x^2+B_y^2+B_z^2$. The rest of the expressions $\left(G_i^k\right)^{(0)}$ and $\left(G_i^k\right)^{(1)}$ are well-known
\citep{Mukhanov,Ruth,Rubakov}:
\be{11b} \left(G_0^0\right)^{(0)}=\frac{3{\mathcal H}^2}{a^2},\quad \left(G_{\beta}^{\alpha}\right)^{(0)}=\left(\frac{2{\mathcal H}'+{\mathcal
H}^2}{a^2}\right)\delta_{\alpha\beta}\, ,\ee
\be{12b} \left(G_0^0\right)^{(1)}=\frac{2}{a^2}\triangle \Phi-\frac{6}{a^2}\mathcal{H}^2\Phi-\frac{6}{a^2}\mathcal{H}\Phi'\, ,\ee
\ba{13b} \left(G_{\beta}^{\alpha}\right)^{(1)}&=&-\frac{2}{a^2}\left[\left(2\mathcal{H}'+\mathcal{H}^2\right)\Phi
+3\mathcal{H}\Phi'+\Phi''\right]\delta_{\alpha\beta}\nn\\
&-&\frac{1}{a^2}\left[\mathcal{H}\left(\frac{\partial B_{\alpha}}{\partial x^{\beta}}+\frac{\partial B_{\beta}}{\partial
x^{\alpha}}\right)+\frac{1}{2}\left(\frac{\partial B_{\alpha}}{\partial x^{\beta}}+\frac{\partial B_{\beta}}{\partial x^{\alpha}}\right)'\,\right]\, .\ea
The expressions $\left(G_i^k\right)^{(2)}$ bear a strong resemblance to $\left(G_i^k\right)^{(1)}$ with due regard for the inequality $\Phi^{(2)}\neq\Psi^{(2)}$
as well as additional contributions from the tensor perturbations:
\be{14b} \left(G_0^0\right)^{(2)}=\frac{2}{a^2}\triangle \Psi^{(2)}-\frac{6}{a^2}\mathcal{H}^2\Phi^{(2)}-\frac{6}{a^2}\mathcal{H}\left(\Psi^{(2)}\right)'\, ,\ee
\ba{15b} &{}& \left(G_{\beta}^{\alpha}\right)^{(2)}=\frac{1}{2a^2}h''_{\alpha\beta} +\frac{1}{a^2}\mathcal{H}h'_{\alpha\beta} -\frac{1}{2a^2}\triangle
h_{\alpha\beta}\nn\\
&-&\frac{2}{a^2}\left[\left(2\mathcal{H}'+\mathcal{H}^2\right)\Phi^{(2)} +\mathcal{H}\left(\Phi^{(2)}\right)'+2\mathcal{H}\left(\Psi^{(2)}\right)'+
\left(\Psi^{(2)}\right)''+\frac{1}{2}\triangle\left(\Phi^{(2)}-\Psi^{(2)}\right)\right]\delta_{\alpha\beta}\nn\\
&+&\frac{1}{a^2}\frac{\partial^2\left(\Phi^{(2)}-\Psi^{(2)}\right)}{\partial x^{\alpha}\partial x^{\beta}}-\frac{1}{a^2}\left[\mathcal{H}\left(\frac{\partial
B_{\alpha}^{(2)}}{\partial x^{\beta}}+\frac{\partial B_{\beta}^{(2)}}{\partial x^{\alpha}}\right)+\frac{1}{2}\left(\frac{\partial B_{\alpha}^{(2)}}{\partial
x^{\beta}}+\frac{\partial B_{\beta}^{(2)}}{\partial x^{\alpha}}\right)'\,\right]\, .\ea
Finally, the quantities $\left(G_0^0\right)^{(\mathrm{II})}$ and $\left(G_{\beta}^{\alpha}\right)^{(\mathrm{II})}$ are quite cumbersome, and it makes no sense to
reproduce them here.

As regards the mixed energy-momentum tensor components in the rhs of \rf{4b}, we resort to the well-known formulas for the analyzed system of point-like particles
\citep{Landau,EZcosm2}:
\be{16b} T^0_0=\frac{\rho c^2}{\sqrt{-g}}\cdot\frac{g_{00}+\tilde v^{\gamma}g_{0\gamma}}{\sqrt{g_{00}+2g_{0\alpha}\tilde v^{\alpha}+g_{\alpha\beta}\tilde
v^{\alpha}\tilde v^{\beta}}}\, ,\ee
\be{17b} T_{\alpha}^0=\frac{\rho c^2}{\sqrt{-g}}\cdot\frac{g_{0\alpha}+\tilde v^{\beta}g_{\alpha\beta}}{\sqrt{g_{00}+2g_{0\mu}\tilde v^{\mu}+ g_{\mu\nu}\tilde
v^{\mu}\tilde v^{\nu}}}\, ,\ee
\be{18b} T_{\beta}^{\alpha}=\frac{\rho c^2}{\sqrt{-g}}\cdot\frac{\tilde v^{\alpha}\left(g_{0\beta}+\tilde v^{\gamma}g_{\gamma\beta}\right)}{\sqrt{g_{00}+
2g_{0\mu}\tilde v^{\mu}+g_{\mu\nu}\tilde v^{\mu}\tilde v^{\nu}}}\, ,\ee
where $g\equiv\mathrm{det}(g_{ik})$. In addition, $\tilde v^{\alpha}$ coincides with $\tilde v_n^{\alpha}$ in the term with the factor $\delta({\bf r}-{\bf
r}_n)$. For instance,
\be{19b} \rho \tilde v^{\alpha}\equiv\sum\limits_nm_n\delta({\bf r}-{\bf r}_n)\tilde v_n^{\alpha}=\sum\limits_n\rho_n\tilde v_n^{\alpha}\, .\ee

With the help of \rf{1b} we immediately write down the metric coefficients $g_{ik}$ up to the second order:
\ba{20b} g_{00}&=&a^2\left(1 + 2\Phi +2\Phi^{(2)}\right),\quad g_{0\alpha}=a^2\left(B_{\alpha} +B_{\alpha}^{(2)}\right)\,
,\nn\\
g_{\alpha\beta}&=&a^2\left(-\delta_{\alpha\beta} +2\Phi\delta_{\alpha\beta} +2\Psi^{(2)}\delta_{\alpha\beta}+h_{\alpha\beta}\right)\, .\ea
Consequently, with the same accuracy
\be{21b} \frac{1}{\sqrt{-g}}=\frac{1}{a^4}\left(1+2\Phi-\frac{1}{2}{\bf B}^2+6\Phi^2-\Phi^{(2)}+3\Psi^{(2)}\right)\, ,\ee
and the formulas \rf{16b}, \rf{17b} and \rf{18b} are reduced to the formulas
\be{22b} T^0_0=\frac{c^2}{a^3} \left(\overline{\rho}+\delta\rho +3\overline{\rho}\Phi +3\delta\rho\Phi+\frac{1}{2}\rho\tilde{v}^2
+\frac{15}{2}\overline{\rho}\Phi^2 -\frac{1}{2}\overline{\rho}{\bf B}^2 +3\overline{\rho}\Psi^{(2)}\right)\, ,\ee
\be{23b} T_{\alpha}^0=\frac{c^2}{a^3} \left(\overline{\rho}B_{\alpha}-\rho\tilde v^{\alpha}+\delta\rho B_{\alpha} +\overline{\rho}B_{\alpha}\Phi+\rho\tilde
v^{\alpha}\Phi +\overline{\rho}B_{\alpha}^{(2)}\right)\, ,\ee
\be{24b} T_{\beta}^{\alpha}= \frac{c^2}{a^3} \left(\rho \tilde v^{\alpha}B_{\beta} -\rho \tilde v^{\alpha}\tilde v^{\beta} \right)\, ,\ee
respectively. Here $\tilde{v}^2\equiv \delta_{\alpha\beta}\tilde v^{\alpha}\tilde v^{\beta}$. By analogy with \cite{Eingorn}, we consider the quantities
$\delta\rho$ and ${\bf \tilde v}_n$ as ``importing'' the first order of smallness in the Einstein equations. In other words, metric corrections, which are
generated by these sources themselves, are assigned the first order. Indeed, in the above-mentioned equations \rf{8a} and \rf{9a} the quantities $\delta\rho$ and
${\bf \tilde v}_n$ (as well as $\Xi$ containing ${\bf \tilde v}_n$) in the rhs play the role of sources generating the first-order metric corrections $\Phi$ and
${\bf B}$ forming the lhs. Therefore, we have omitted such terms as, for example, $\sim\delta\rho\Phi^2$ in \rf{22b} since this term is much smaller than the
summand $\sim\delta\rho\Phi$  in the same parentheses at all cosmological scales and would import the third order of smallness (see also \cite{Chisari} for the
similar reasoning). The established thorough separation of the first- and second-order summands in the Einstein equations is strongly corroborated in
subsection~\ref{subsec34} below.

Once again we employ a helpful decomposition
\be{25b} T_i^k=\left(T_i^k\right)^{(0)}+\left(T_i^k\right)^{(1)}+\left(T_i^k\right)^{(2)}+\left(T_i^k\right)^{(\mathrm{II})}\, .\ee
The only nonzero component with the superscript ``(0)'' is $\left(T_0^0\right)^{(0)}=\overline{\rho}c^2/a^3$. The components $\left(T_i^k\right)^{(1)}$ (or, by
analogy, $\left(T_i^k\right)^{(2)}$) are constructed from the terms being linear in the quantities $\delta\rho$, ${\bf\tilde v}$, $\Phi$, ${\bf B}$ (or
$\Psi^{(2)}$, ${\bf B}^{(2)}$):
\be{26b} \left(T_0^0\right)^{(1)}= \frac{c^2}{a^3}\delta\rho +\frac{3\overline{\rho}c^2}{a^3}\Phi,\quad \left(T^{0}_{\alpha}\right)^{(1)}=
\frac{\overline{\rho}c^2}{a^3}B_{\alpha}-\frac{c^2}{a^3}\rho\tilde v^{\alpha}, \quad \left(T^{\alpha}_{\beta}\right)^{(1)}= 0\, ,\ee
\be{27b} \left(T^0_0\right)^{(2)}= \frac{3\overline{\rho}c^2}{a^3}\Psi^{(2)},\quad \left(T^{0}_{\alpha}\right)^{(2)}=
\frac{\overline{\rho}c^2}{a^3}B_{\alpha}^{(2)}, \quad \left(T^{\alpha}_{\beta}\right)^{(2)}= 0\, .\ee
Finally, the components with the superscript ``(II)'' contain products of $\delta\rho$, ${\bf\tilde v}$, $\Phi$ or ${\bf B}$:
\ba{28b} \left(T^0_0\right)^{(\mathrm{II})}&=& \frac{3c^2}{a^3}\delta\rho\Phi+\frac{c^2}{2a^3}\rho\tilde{v}^2 +\frac{15\overline{\rho}c^2}{2a^3}\Phi^2
 - \frac{\overline{\rho}c^2}{2a^3}{\bf B}^2\, ,\nn\\
\left(T^{0}_{\alpha}\right)^{(\mathrm{II})}&=& \frac{c^2}{a^3}\delta\rho B_{\alpha} +\frac{\overline{\rho}c^2}{a^3}\Phi B_{\alpha} +\frac{c^2}{a^3}\rho\tilde
v^{\alpha}\Phi\, ,\\
\left(T^{\alpha}_{\beta}\right)^{(\mathrm{II})}&=& \frac{c^2}{a^3}\rho\tilde v^{\alpha}B_{\beta} - \frac{c^2}{a^3}\rho\tilde v^{\alpha}\tilde v^{\beta}\nn\, .\ea
It is worth mentioning that owing to the zero value of $\left(T^{\alpha}_{\beta}\right)^{(1)}$ the anisotropic stress vanishes in the first-order approximation.
This is the cogent reason for using the same designation $\Phi\equiv\Phi^{(1)}=\Psi^{(1)}$ for the equal first-order scalar perturbations $\Phi^{(1)}$ and
$\Psi^{(1)}$ from the very outset. Nevertheless, owing to the nonzero values of $\left(G^{\alpha}_{\beta}\right)^{(\mathrm{II})}$ and
$\left(T^{\alpha}_{\beta}\right)^{(\mathrm{II})}$ (or, more precisely, their combinations $Q_{\alpha\beta}$ introduced below) the anisotropic stress does not
vanish in the second-order approximation. Thus, generally speaking, the second-order scalar perturbations $\Phi^{(2)}$ and $\Psi^{(2)}$ are unequal:
$\Phi^{(2)}\neq\Psi^{(2)}$.

Let us conclude this subsection by introducing the promised helpful combinations
\be{29b} Q_{ik}\equiv\kappa\left(T^{k}_{i}\right)^{(\mathrm{II})} -\left(G^{k}_{i}\right)^{(\mathrm{II})} \ee
and presenting their explicit expressions without concealing anything, even despite the quite cumbersome form of some of them:
\ba{30b} Q_{00} &=& \frac{\kappa c^2}{2a^3}\rho\tilde{v}^2-\frac{3\kappa^2 c^4}{4a^4}\Xi^2+\frac{6\kappa c^2}{a^3}\mathcal{H}\Xi\Phi
-\left(\frac{3\kappa\overline{\rho}c^2}{2a^3}
+\frac{15}{a^2}\mathcal{H}^2\right)\Phi^2\nn\\
&+&\left(\frac{\kappa\overline{\rho} c^2}{2a^3} +\frac{3}{a^2}\mathcal{H}^2\right){\bf B}^2-\frac{2}{a^2}\Phi\triangle\Phi-
\frac{3}{a^2}\left(\nabla\Phi\right)^2+\frac{2}{a^2}\mathcal{H}{\bf B}\nabla\Phi\nn\\
&-&\frac{1}{4a^2}{\bf B}\triangle{\bf B} +\frac{1}{8a^2}\triangle\left({\bf B}^2\right) +\frac{1}{4a^2}\nabla\left[\left({\bf B}\nabla\right){\bf
B}\right]-\frac{\kappa c^2}{a^3}\rho {\bf \tilde v}{\bf B}\, ,\ea
\ba{31b} Q_{0\beta}&=& \frac{\kappa c^2}{a^3}\delta\rho B_{\beta}+ \frac{\kappa \overline{\rho}c^2}{a^3}\Phi B_{\beta} +\frac{\kappa c^2}{a^3}\rho\tilde
v^{\beta}\Phi-\frac{1}{a^2}B_{\beta}\triangle \Phi -
\frac{1}{a^2}\mathcal{H}\frac{\partial \left({\bf B}^2\right)}{\partial x^{\beta}} \nn \\
&+&\frac{5}{a^2}\mathcal{H}\frac{\partial \left(\Phi^2\right)}{\partial x^{\beta}} + \frac{\kappa c^2}{a^3}\Xi\frac{\partial \Phi}{\partial x^{\beta}}+
\frac{1}{a^2}\delta^{\alpha\gamma}\frac{\partial B_{\gamma}}{\partial x^{\beta}}\frac{\partial \Phi}{\partial
x^{\alpha}}-\frac{1}{a^2}\delta^{\alpha\gamma}B_{\gamma}\frac{\partial^2 \Phi}{\partial x^{\alpha}\partial x^{\beta}}\, ,\ea
\ba{32b} Q_{11}&=&-\frac{\kappa c^2}{a^3}\rho\tilde v_{x}^2-\frac{\kappa^2 c^4}{4a^4}\Xi^2+\frac{3\kappa
c^2}{a^3}\mathcal{H}\Xi\Phi+\left(\frac{4\kappa\overline{\rho} c^2}{a^3}-\frac{5}{a^2}\mathcal{H}^2\right)\Phi^2-\frac{1}{a^2}\mathcal{H}^2{\bf B}^2\nn\\
&-&\frac{4}{a^2}\Phi\triangle\Phi +\frac{4}{a^2}\Phi\frac{\partial^2 \Phi}{\partial x^2}-\frac{3}{a^2}\left(\nabla\Phi\right)^2 +
\frac{2}{a^2}\left(\frac{\partial \Phi}{\partial x}\right)^2-\frac{2}{a^2}\mathcal{H}B_{x}\frac{\partial \Phi}{\partial x}-\frac{3}{4a^2}{\bf B}\triangle{\bf
B}\nn\\
&+&\frac{1}{2a^2}B_{x}\triangle B_{x}+\frac{1}{8a^2}\triangle\left({\bf B}^2\right) -\frac{1}{4a^2}\nabla\left[\left({\bf B}\nabla\right){\bf
B}\right]-\frac{\kappa c^2}{a^3}\rho{\bf\tilde v}{\bf B}+\frac{\kappa c^2}{a^3}\rho\tilde v_{x}B_{x}\nn \\
&+&\frac{2}{a^2}\mathcal{H}\Phi\frac{\partial B_{x}}{\partial x}+\frac{\kappa c^2}{a^3}\Xi\frac{\partial B_{x}}{\partial x}-\frac{1}{a^2}\frac{\partial
B_{z}}{\partial z}\frac{\partial B_{y}}{\partial y} + \frac{1}{2a^2}\left(\frac{\partial B_{y}}{\partial z}\right)^2 +\frac{1}{2a^2}\left(\frac{\partial
B_{z}}{\partial y}\right)^2 \nn \\
&+&\frac{1}{a^2}B_{z}\frac{\partial^2 B_{z}}{\partial y^2} +\frac{1}{a^2}B_{y}\frac{\partial^2 B_{y}}{\partial z^2}-\frac{1}{a^2}B_{y}\frac{\partial^2
B_{z}}{\partial y\partial z} -\frac{1}{a^2}B_{z}\frac{\partial^2 B_{y}}{\partial y\partial z}\, ,\ea
$Q_{22}$ and $Q_{33}$ are analogous,
\ba{33b} Q_{12}&=& \frac{\kappa c^2}{2a^3}\rho\tilde v_{y}B_{x}-\frac{\kappa c^2}{2a^3}\rho\tilde v_{y}\tilde v_{x} -\frac{1}{4a^2}B_y\frac{\partial^2
B_x}{\partial z^2} + \frac{1}{4a^2}B_y\frac{\partial^2 B_x}{\partial y^2}-
\frac{1}{4a^2}B_y\frac{\partial^2 B_x}{\partial x^2} -\frac{1}{2a^2}B_{y}\frac{\partial^2 B_{y}}{\partial x\partial y}  \nn \\
&+& \frac{1}{2a^2}B_{z}\frac{\partial^2 B_{y}}{\partial x\partial z}-\frac{1}{2a^2}B_{z}\frac{\partial^2 B_{z}}{\partial x\partial y}-\frac{1}{4a^2}\frac{\partial
B_{x}}{\partial z}\frac{\partial B_{y}}{\partial z} +\frac{1}{2a^2}\frac{\partial B_{z}}{\partial z}\frac{\partial B_{y}}{\partial x}-
\frac{1}{4a^2}\frac{\partial B_{z}}{\partial y}\frac{\partial B_{z}}{\partial x}\nn\\
&+& \frac{1}{a^2}\mathcal{H}\Phi\frac{\partial B_{x}}{\partial y}+\frac{\kappa c^2}{2a^3}\Xi\frac{\partial B_{x}}{\partial
y}-\frac{1}{a^2}\mathcal{H}B_{x}\frac{\partial \Phi}{\partial y}+\frac{1}{a^2}\frac{\partial \Phi}{\partial y}\frac{\partial \Phi}{\partial x}
+\frac{2}{a^2}\Phi\frac{\partial^2 \Phi}{\partial x\partial y}+\{x\leftrightarrow y\}\, ,\ea
where $\{x\leftrightarrow y\}$ stands for exactly the same terms with the occurring everywhere replacement of $x$ by $y$ and vice versa, $Q_{13}$ and $Q_{23}$ are
analogous. The enumerated formulas have been derived, in particular, through the instrumentality of the formulas \rf{10b}, \rf{28b} and equations from
Section~\ref{sec2} including the following direct consequence of \rf{2a}: $\mathcal{H}^2-\mathcal H'=\kappa
a^2\overline{\varepsilon}/2=\kappa\overline{\rho}c^2/(2a)$. From \rf{31b} and \rf{32b} we get
\ba{34b} \frac{\partial Q_{0\beta}}{\partial x^{\beta}}&\equiv&\frac{\partial Q_{01}}{\partial x}+\frac{\partial Q_{02}}{\partial y}+\frac{\partial
Q_{03}}{\partial z}= \triangle\left(\frac{\kappa c^2}{a^3}\Xi\Phi- \frac{1}{a^2}\mathcal{H}{\bf
B}^2+\frac{5}{a^2}\mathcal{H}\Phi^2\right)\nn\\
&-&\frac{\kappa c^2}{a^3}\rho{\bf\tilde v}\left(\nabla\Phi\right) +\frac{\kappa c^2}{a^3}\nabla\Phi\nabla\Xi+\frac{3\kappa c^2}{a^3}\mathcal{H}{\bf
B}(\nabla\Xi)\, ,\ea
\ba{35b} Q_{\alpha\alpha}&\equiv& Q_{11}+Q_{22}+Q_{33}= - \frac{\kappa c^2}{a^3}\rho \tilde v^2-\frac{3\kappa^2 c^4}{4a^4}\Xi^2+\frac{9\kappa
c^2}{a^3}\mathcal{H}\Xi\Phi \nn \\
&+&\left(\frac{12\kappa\overline{\rho} c^2}{a^3}-\frac{15}{a^2}\mathcal{H}^2\right)\Phi^2-\frac{3}{a^2}\mathcal{H}^2{\bf
B}^2-\frac{8}{a^2}\Phi\triangle\Phi-\frac{7}{a^2}\left(\nabla\Phi\right)^2-\frac{2}{a^2}\mathcal{H}{\bf B}\nabla\Phi\nn\\
&-&\frac{5}{4a^2}{\bf B}\triangle{\bf B} +\frac{5}{8a^2}\triangle\left({\bf B}^2\right) -\frac{3}{4a^2}\nabla\left[\left({\bf B}\nabla\right){\bf
B}\right]-\frac{2\kappa c^2}{a^3}\rho{\bf \tilde v}{\bf B}\, ,\ea
respectively. Hereinafter summation over repeated Greek subscripts is implied without superfluous decoding. It should be also noted that naturally
$Q_{ik}=Q_{ki}$, in complete agreement with the symmetry inherent in the Einstein equations, which we write down in the very next subsection.

\

\subsection{Scalar, Vector and Tensor Sectors}\label{subsec32}

Substituting \rf{5b} and \rf{25b} into \rf{4b} with due account taken of \rf{29b}, we immediately get:

\

\ $\bullet$ \ $00$-component:
\ba{36b} &{}& \frac{3\mathcal{H}^2}{a^2} +\left(\frac{2}{a^2}\triangle\Phi-\frac{6}{a^2}\mathcal{H}^2\Phi-\frac{6}{a^2}\mathcal{H}\Phi'\right)
+\left(\frac{2}{a^2}\triangle \Psi^{(2)}-\frac{6}{a^2}\mathcal{H}^2\Phi^{(2)}-\frac{6}{a^2}\mathcal{H}\left(\Psi^{(2)}\right)'\right) \nn \\
&=& \frac{\kappa \overline{\rho}c^2}{a^3} +\Lambda +\left(\frac{\kappa c^2}{a^3}\delta\rho +\frac{3\kappa\overline{\rho} c^2}{a^3}\Phi\right) +
\frac{3\kappa\overline{\rho}c^2}{a^3}\Psi^{(2)} + Q_{00}\, ;\ea

\

\ $\bullet$ \ $0\beta$-components:
\ba{37b} &{}& \left(\frac{1}{2a^2}\triangle B_{\beta} +\frac{2}{a^2}\mathcal H\frac{\partial \Phi}{\partial x^{\beta}} +\frac{2}{a^2}\frac{\partial
\Phi'}{\partial x^{\beta}}\right) +\left(\frac{1}{2a^2}\triangle B_{\beta}^{(2)} +\frac{2}{a^2}\mathcal H\frac{\partial \Phi^{(2)}}{\partial x^{\beta}}+
\frac{2}{a^2}\frac{\partial \left(\Psi^{(2)}\right)'}{\partial x^{\beta}}\right) \nn \\
&=& \left(\frac{\kappa\overline{\rho} c^2}{a^3}B_{\beta}-\frac{\kappa c^2}{a^3}\rho\tilde v^{\beta} \right) +\frac{\kappa\overline{\rho}c^2}{a^3}B_{\beta}^{(2)}
+Q_{0\beta}\, ;\ea

\

\ $\bullet$ \ $11$-component:
\ba{38b} &{}& \frac{2\mathcal{H}'+\mathcal{H}^2}{a^2} -\left(\frac{2}{a^2}\left(2\mathcal{H}'+\mathcal{H}^2\right)\Phi + \frac{6}{a^2}\mathcal{H}\Phi'
+\frac{2}{a^2}\Phi'' +\frac{2}{a^2}\mathcal{H}\frac{\partial B_{x}}{\partial x} +
\frac{1}{a^2}\frac{\partial B_{x}'}{\partial x}\right) \nn \\
&-& \left(\frac{2}{a^2}\left(2\mathcal{H}'+\mathcal{H}^2\right)\Phi^{(2)}+\frac{2}{a^2}\mathcal{H}\left(\Phi^{(2)}\right)'+
\frac{4}{a^2}\mathcal{H}\left(\Psi^{(2)}\right)'+\frac{2}{a^2}\left(\Psi^{(2)}\right)''\right.\nn\\
&+&\left.\frac{2}{a^2}\mathcal{H}\frac{\partial B_{x}^{(2)}}{\partial x}+\frac{1}{a^2}\frac{\partial}{\partial
x}\left(B_{x}^{(2)}\right)'-\frac{1}{a^2}\frac{\partial^2\Phi^{(2)}}{\partial x^2}+\frac{1}{a^2}\frac{\partial^2\Psi^{(2)}}{\partial x^2}+
\frac{1}{a^2}\triangle\Phi^{(2)}-\frac{1}{a^2}\triangle\Psi^{(2)}\right)\nn\\
&+&\frac{1}{2a^2}h''_{11} +\frac{1}{a^2}\mathcal{H}h'_{11} -\frac{1}{2a^2}\triangle h_{11} = \Lambda +Q_{11}\, ,\ea
$22$- and $33$-components are similar;

\

\ $\bullet$ \ $12$-component:
\ba{39b} &-&\left(\frac{1}{a^2}\mathcal{H}\frac{\partial B_{x}}{\partial y} +\frac{1}{2a^2}\frac{\partial B_{x}'}{\partial y}
+\frac{1}{a^2}\mathcal{H}\frac{\partial B_{y}}{\partial x} +\frac{1}{2a^2}\frac{\partial B_{y}'}{\partial x}\right)\nn\\
&-&\left(\frac{1}{a^2}\mathcal{H}\frac{\partial B_{x}^{(2)}}{\partial y}+\frac{1}{2a^2}\frac{\partial }{\partial y}\left(B_{x}^{(2)}\right)'+
\frac{1}{a^2}\mathcal{H}\frac{\partial B_{y}^{(2)}}{\partial x}+\frac{1}{2a^2}\frac{\partial }{\partial
x}\left(B_{y}^{(2)}\right)'-\frac{1}{a^2}\frac{\partial^2\Phi^{(2)}}{\partial x\partial y}+ \frac{1}{a^2}\frac{\partial^2\Psi^{(2)}}{\partial x\partial
y}\right)\nn\\
&+&\frac{1}{2a^2}h''_{12} +\frac{1}{a^2}\mathcal{H}h'_{12} -\frac{1}{2a^2}\triangle h_{12} = Q_{12}\, ,\ea
$13$- and $23$-components are similar.

\

All these equations clearly demonstrate that the second-order scalar, vector and tensor perturbations (represented by $\Phi^{(2)}$, $\Psi^{(2)}$; ${\bf B}^{(2)}$;
$h_{\alpha\beta}$, respectively) do not mix \citep{Baumann} and are generated, in particular, by the quadratic combinations of the first-order scalar and vector
perturbations $\Phi$; ${\bf B}$.

Now it is just the right time for the standard ``scalar--vector--tensor'' decomposition of $Q_{\alpha\beta}$:
\be{40b} Q_{\alpha\beta}=Q^{(0)}\delta_{\alpha\beta}+\frac{\partial^2 Q^{(\mathrm{S})}}{\partial x^{\alpha}\partial x^{\beta}}+\frac{\partial
Q^{(\mathrm{V})}_{\alpha}}{\partial x^{\beta}}+\frac{\partial Q^{(\mathrm{V})}_{\beta}}{\partial x^{\alpha}}+Q^{(\mathrm{T})}_{\alpha\beta}\, ,\ee
where $Q^{(0)}$ and $Q^{(\mathrm{S})}$ describe the scalar sector while ${\bf Q}^{(\mathrm{V})}$ and $Q^{(\mathrm{T})}_{\alpha\beta}$ describe the vector and
tensor sectors, respectively, and satisfy the corresponding conditions:
\be{41b} \frac{\partial Q^{(\mathrm{V})}_{\alpha}}{\partial x^{\alpha}}\equiv0,\quad \frac{\partial Q^{(\mathrm{T})}_{\alpha\beta}}{\partial
x^{\alpha}}\equiv0,\quad Q^{(\mathrm{T})}_{\alpha\alpha}\equiv0\, .\ee
According to \rf{40b} and \rf{41b}, the introduced functions $Q^{(0)}$, $Q^{(\mathrm{S})}$ and ${\bf Q}^{(\mathrm{V})}$ can be determined as solutions of the
corresponding equations
\be{42b} \triangle Q^{(0)}=\frac{1}{2}\triangle Q_{\alpha\alpha}-\frac{1}{2}\frac{\partial^2Q_{\alpha\beta}}{\partial x^{\alpha}\partial x^{\beta}},\quad
\triangle\triangle Q^{(\mathrm{S})}=-\frac{1}{2}\triangle Q_{\alpha\alpha}+\frac{3}{2}\frac{\partial^2Q_{\alpha\beta}}{\partial x^{\alpha}\partial x^{\beta}}\, ,
\ee
\be{43b} \triangle\triangle Q_{\beta}^{(\mathrm{V})}=\triangle\frac{\partial Q_{\alpha\beta}}{\partial x^{\alpha}}-\frac{\partial^3Q_{\alpha\gamma}}{\partial
x^{\beta}\partial x^{\alpha}\partial x^{\gamma}}\, ,\ee
then the remaining unknown functions $Q^{(\mathrm{T})}_{\alpha\beta}$ can be easily found from \rf{40b}. They act as the sole sources of gravitational waves (see
the rhs of the equation \rf{47b} for $h_{\alpha\beta}$ below).

Let us synchronously perform the standard ``scalar--vector'' decomposition of $Q_{0\alpha}$:
\be{44b} Q_{0\alpha}=\frac{\partial Q^{(\parallel)}}{\partial x^{\alpha}}+Q^{(\perp)}_{\alpha},\quad \frac{\partial Q^{(\perp)}_{\alpha}}{\partial
x^{\alpha}}\equiv0\, ,\ee
where $Q^{(\parallel)}$ and ${\bf Q}^{(\perp)}$ denote the scalar and vector contributions, respectively. According to \rf{44b}, they can be determined as
solutions of the corresponding equations
\be{45b} \triangle Q^{(\parallel)}=\frac{\partial Q_{0\alpha}}{\partial x^{\alpha}},\quad \triangle Q_{\alpha}^{(\perp)}=\triangle
Q_{0\alpha}-\frac{\partial^2Q_{0\beta}}{\partial x^{\alpha}\partial x^{\beta}}\, .\ee

Finally, in order to abridge the notation, let us introduce the handy mixed-order quantities
\be{46b} \Phi^{(12)}\equiv\Phi+\Phi^{(2)},\quad \Psi^{(12)}\equiv\Phi+\Psi^{(2)},\quad {\bf B}^{(12)}\equiv {\bf B}+{\bf B}^{(2)}\, .\ee

Then the Einstein equations \rf{36b}--\rf{39b} can be eventually rewritten as follows:

\

\ $\bullet$ \ tensor sector:
\be{47b} h''_{\alpha\beta} +2\mathcal{H}h'_{\alpha\beta} -\triangle h_{\alpha\beta} = 2a^2 Q^{(\mathrm{T})}_{\alpha\beta}\, ;\ee

\

\ $\bullet$ \ vector sector:
\be{48b} \triangle {\bf B}^{(12)} -\frac{2\kappa\overline{\rho}c^2}{a}{\bf B}^{(12)} =-\frac{2\kappa c^2}{a}\left(\rho {\bf \tilde v} -\nabla\Xi\right) +2a^2 {\bf
Q}^{(\perp)}\, ,\ee
\be{49b} \left({\bf B}^{(12)}\right)' +2\mathcal{H}{\bf B}^{(12)} =-2a^2 {\bf Q}^{(\mathrm{V})}\, ;\ee

\

\ $\bullet$ \ scalar sector:
\be{50b} \Phi^{(12)}-\Psi^{(12)} =a^2 Q^{(\mathrm{S})}\, ,\ee
\be{51b} \triangle \Psi^{(12)} -\frac{3\kappa\overline{\rho}c^2}{2a}\Psi^{(12)} -3\mathcal{H}\left[\left(\Psi^{(12)}\right)'+\mathcal{H}\Phi^{(12)}\right]
=\frac{\kappa c^2}{2a}\delta\rho + \frac{a^2}{2}Q_{00}\, ,\ee
\be{52b} \left(\Psi^{(12)}\right)' +\mathcal H\Phi^{(12)} =-\frac{\kappa c^2}{2a} \Xi +\frac{a^2}{2}Q^{(\parallel)}\, ,\ee
\be{53b} \triangle\Psi^{(12)}-\triangle\Phi^{(12)} -2\left(\Psi^{(12)}\right)'' -4\mathcal{H}\left(\Psi^{(12)}\right)' -2\mathcal{H}\left(\Phi^{(12)}\right)'
-2\left(2\mathcal{H}'+\mathcal{H}^2\right)\Phi^{(12)} = a^2 Q^{(0)}\, .\ee
Substituting \rf{52b} into \rf{51b}, we get
\be{54b} \triangle \Psi^{(12)} -\frac{3\kappa\overline{\rho}c^2}{2a}\Psi^{(12)} =\frac{\kappa c^2}{2a}\delta\rho -\frac{3\kappa c^2}{2a}\mathcal{H}\Xi
+\frac{a^2}{2}Q_{00} +\frac{3a^2}{2}\mathcal{H}Q^{(\parallel)}\, .\ee
Recalling \rf{46b} along with \rf{8a} and \rf{9a}, we reduce \rf{48b}, \rf{50b} and \rf{54b} to the equations
\be{55b} \triangle {\bf B}^{(2)} -\frac{2\kappa\overline{\rho}c^2}{a}{\bf B}^{(2)} = 2a^2 {\bf Q}^{(\perp)}\, ,\ee
\be{56b} \Phi^{(2)}-\Psi^{(2)} =a^2 Q^{(\mathrm{S})},\quad \triangle \Psi^{(2)} -\frac{3\kappa\overline{\rho}c^2}{2a}\Psi^{(2)} = \frac{a^2}{2}Q_{00}
+\frac{3a^2}{2}\mathcal{H}Q^{(\parallel)}\, .\ee

Thus, we have derived the ``master'' equations \rf{47b}, \rf{55b} and \rf{56b} for the sought-for second-order cosmological perturbations. In the next subsection
we show that the remaining ``non-master'' equations \rf{49b}, \rf{52b} and \rf{53b} are satisfied automatically provided that one takes advantage of the equations
of motion governing the particle dynamics.

\

\subsection{Verification of Equations}\label{subsec33}

Without going into detail, let us outline the proof that the scalar sector equations \rf{52b} and \rf{53b} containing temporal derivatives are really satisfied.
In the first place, one finds a derivative of \rf{54b} with respect to $\eta$ and further expresses $\left(\Psi^{(12)}\right)'$ from \rf{52b}. As a result,
\ba{57b} \frac{3\kappa c^2}{2a}\mathcal{H}\Xi' &=& \mathcal{H}\triangle\Phi^{(12)}-\frac{3\kappa\overline{\rho}c^2}{2a}\mathcal{H}\Phi^{(12)}-
\frac{3\kappa\overline{\rho}c^2}{2a}\mathcal{H}\Psi^{(12)}-\frac{\kappa c^2}{2a}\mathcal{H}\delta\rho\nn\\
&-&\frac{a^2}{2}\triangle Q^{(\parallel)}+\frac{3\kappa\overline{\rho}c^2 a}{4}Q^{(\parallel)}
+\frac{3}{2}\left(a^2\mathcal{H}Q^{(\parallel)}\right)'+\frac{1}{2}\left(a^2Q_{00}\right)'\, .\ea
In the second place, one substitutes the expression for $\left(\Psi^{(12)}\right)'$ from \rf{52b} into \rf{53b}. As a result,
\ba{58b} \frac{3\kappa c^2}{2a}\mathcal{H}\Xi' &=&
\frac{3}{2}\mathcal{H}\triangle\Phi^{(12)}-\frac{3}{2}\mathcal{H}\triangle\Psi^{(12)}-3\mathcal{H}\left(\mathcal{H}^2-\mathcal{H}'\right)\Phi^{(12)}-\frac{3\kappa
c^2}{2a}\mathcal{H}^2\Xi \nn\\
&+&  3a^2\mathcal{H}^2Q^{(\parallel)} +\frac{3}{2}\mathcal{H}\left(a^2Q^{(\parallel)}\right)' + \frac{3a^2}{2}\mathcal{H}Q^{(0)}\, .\ea
Therefore, it is enough to show that the rhs of \rf{57b} is really equal to the rhs of \rf{58b} since the lhs of these equations is the same, and then to prove
either equation. We have successfully coped with both these onerous tasks. Equating the right-hand sides, after lengthy calculation one eventually arrives at an
identity. The following auxiliary formulas should be used on the way:
\be{59b} \rho{\bf\tilde v}'\equiv\sum\limits_{n}m_n\delta({\bf r}-{\bf r}_n){\bf \tilde{v}}'_n =-\mathcal{H}\rho{\bf\tilde
v}-\overline{\rho}\nabla\Phi-\overline{\rho}\mathcal{H}{\bf B} \ee
in the first-order approximation \citep{Eingorn};
\ba{60b} \left(\rho\tilde{v}^2\right)'&\equiv&\left(\sum\limits_{n}m_n\delta({\bf r}-{\bf r}_n)\tilde{v}_n^2\right)' =
2\sum\limits_{n}m_n\delta({\bf r}-{\bf r}_n){\bf \tilde v}_n{\bf \tilde v}'_n \nn \\
&=&-2\mathcal{H}\rho{\tilde v}^2 -2\rho{\bf \tilde v}\left(\nabla\Phi \right) -2\mathcal{H}\rho{\bf \tilde v}{\bf B} \ea
within the adopted accuracy. The underlying equations of motion of the $n$-th particle have the form ${\bf\tilde v}'_n =-\mathcal{H}{\bf\tilde
v}_n-\nabla\Phi-\mathcal{H}{\bf B}$ \citep{Eingorn}.

When the desired identity is achieved, it is enough to prove, for instance, the correctness of \rf{58b}. Now the accuracy of \rf{59b} is insufficient, and it is
necessary to take advantage of the spacetime interval for the $n$-th particle:
\ba{61b} ds_n&=&a\left\{1 + 2\Phi+2\Phi^{(2)} + 2\left(B_{\alpha}+B_{\alpha}^{(2)}\right) \tilde{v}_n^{\alpha}\right. \nn \\
&+&\left.\left[\left(-1+ 2\Phi+2\Psi^{(2)}\right)\delta_{\alpha\beta}+h_{\alpha\beta}\right]\tilde{v}_n^{\alpha} \tilde{v}_n^{\beta}\right\}^{1/2}d\eta\, , \ea
where the metric corrections are computed at the point ${\bf r}={\bf r}_n$ and, as usual, do not include the divergent contributions from the considered particle
itself. For the sake of simplicity we can confine ourselves to those terms in \rf{58b} which are not quadratic in particle velocities, then the Lagrange equations
of motion have the form
\ba{62b} {\bf\tilde v}'_n&=&-\mathcal{H}{\bf\tilde v}_n-\nabla\Phi+\mathcal{H}{\bf B}+{\bf B}'-3\mathcal{H}{\bf\tilde v}_n\Phi\nn\\
&-&\nabla\left(\Phi^2\right)-\nabla\Phi^{(2)} -\mathcal{H}\Phi{\bf B} +\mathcal{H}{\bf B}^{(2)} +\left({\bf B}^{(2)}\right)'\, . \ea
Multiplication of \rf{62b} by $\rho_n$ with subsequent summation over $n$ gives
\ba{63b} \rho{\bf\tilde v}' &=& -\mathcal{H}\rho{\bf\tilde v}-\overline{\rho}\nabla\Phi -\delta\rho\nabla\Phi+ \overline{\rho}\mathcal{H}{\bf
B}-\mathcal{H}\delta\rho{\bf B} +\overline{\rho}{\bf B}'-3\mathcal{H}\rho{\bf\tilde v}\Phi\nn\\
&-& \overline{\rho}\nabla\left(\Phi^2\right)  - \overline{\rho}\nabla\Phi^{(2)} - \overline{\rho}\mathcal{H}\Phi{\bf B} +\overline{\rho}\mathcal{H}{\bf B}^{(2)} +
\overline{\rho}\left({\bf B}^{(2)}\right)'\, , \ea
where the last equation of the triplet \rf{11a} has been used to replace the summand $\delta\rho{\bf B}'$ by $-2\mathcal{H}\delta\rho{\bf B}$. We have also
dropped all terms which would import the third order of smallness in the Einstein equations. If one additionally omits the terms importing the second order, then
\rf{63b} is reduced exactly to \rf{59b}.

Being armed with \rf{63b}, after exhausting calculation one turns \rf{58b} into an identity. Thus, both initial non-master scalar sector equations \rf{52b} and
\rf{53b} are satisfied. The same applies to \rf{49b}. Indeed, suffice it to demonstrate that
\be{64b} \triangle\left[\left({\bf B}^{(12)}\right)' +2\mathcal{H}{\bf B}^{(12)} +2a^2 {\bf Q}^{(\mathrm{V})}\right] -\frac{2\kappa\overline{\rho}
c^2}{a}\left[\left({\bf B}^{(12)}\right)' +2\mathcal{H}{\bf B}^{(12)} +2a^2 {\bf Q}^{(\mathrm{V})}\right] =0\, .\ee
Recalling \rf{48b}, one can reduce \rf{64b} to the following equation:
\ba{65b} &-&\frac{2\kappa\overline{\rho}c^2}{a}\mathcal{H}{\bf B}^{(12)} -\frac{2\kappa c^2}{a}\mathcal{H}\left(\rho{\bf\tilde v} -\nabla\Xi\right) -
\frac{2\kappa c^2}{a}\left(\rho{\bf\tilde v} -\nabla\Xi\right)'\nn\\
&+&8a^2\mathcal{H} {\bf Q}^{(\perp)}+2a^2 \left({\bf Q}^{(\perp)}\right)' +2a^2\triangle{\bf Q}^{(\mathrm{V})} -4\kappa\overline{\rho}c^2a {\bf Q}^{(\mathrm{V})}
=0\, .\ea
In the framework of the above-mentioned simplification, that is without products of velocities, substitution of \rf{58b} and \rf{63b} into \rf{65b} eventually
leads to the desired identity. Thus, the initial non-master vector sector equation \rf{49b} is satisfied as well. Obviously, the same applies to the gauge
conditions \rf{2b} and \rf{3b} since exactly the same gauge conditions hold true for the corresponding right-hand sides of \rf{55b} and \rf{47b}.

\

\subsection{Self-consistent Separation of Summands}\label{subsec34}

In the previous subsection we have demonstrated that the functions $\Phi^{(2)}$, $\Psi^{(2)}$; ${\bf B}^{(2)}$; $h_{\alpha\beta}$ determined as solutions of
\rf{56b}, \rf{55b} and \rf{47b}, respectively, satisfy all Einstein equations in the second-order approximation. The following relevant question arises: is the
undertaken separation of the first- and second-order terms well-grounded and self-consistent? In other words, do we correctly and logically assign orders to
different summands?

Of course, the answer is affirmative. As an illustrative example, let us single out two types of terms in \rf{51b}, namely, those which are either present, or
absent in the corresponding equation in the framework of the first-order approximation
\be{66b} \triangle \Phi -\frac{3\kappa\overline{\rho}c^2}{2a}\Phi -3\mathcal{H}\left(\Phi'+\mathcal{H}\Phi\right)+\mathcal{H}\nabla{\bf B} =\frac{\kappa
c^2}{2a}\delta\rho\, ,\ee
which is equivalent to \rf{8a} in view of the gauge condition \rf{4a} and the first equation of the triplet \rf{11a}. The vanishing last term in the lhs of
\rf{66b} is momentarily reinstated since it can be considered as being initially present in the corresponding $00$-component of the Einstein equations as a part
of $\left(G_0^0\right)^{(1)}$ before applying the gauge condition \rf{4a}.

We designate the first derivatives with respect to each comoving spatial coordinate and conformal time as $1/L$ and $1/\Upsilon$, respectively, as well as ascribe
the orders of smallness $\epsilon$ and $\epsilon^2$ to the first- and second-order metric corrections. Then, for instance, $\triangle\Phi\sim\epsilon/L^2$ while
$\left(\Psi^{(2)}\right)'\sim\epsilon^2/\Upsilon$. As a result, taking into account the explicit expression \rf{30b} for $Q_{00}$, we have $6$ terms of the first
type (present in \rf{51b} as well as in \rf{66b}), namely,
\be{67b} \frac{1}{L^2}\epsilon,\quad \frac{\kappa\overline{\rho}c^2}{a}\epsilon,\quad \frac{\mathcal{H}}{\Upsilon}\epsilon,\quad \mathcal{H}^2\epsilon,\quad
\frac{\mathcal{H}}{L}\epsilon,\quad \frac{\kappa c^2}{a}\delta\rho\, ,\ee
and $9$ terms of the second type (present in \rf{51b}, but absent in \rf{66b}):
\be{68b} \frac{1}{L^2}\epsilon^2,\quad \frac{\kappa\overline{\rho}c^2}{a}\epsilon^2,\quad \frac{\mathcal{H}}{\Upsilon}\epsilon^2,\quad
\mathcal{H}^2\epsilon^2,\quad \frac{\mathcal{H}}{L}\epsilon^2,\quad \frac{\kappa c^2}{a}\rho\tilde{v}^2,\quad \frac{\kappa c^2\mathcal{H}}{a}\Xi\epsilon,\quad
\frac{\kappa c^2}{a}\rho{\bf \tilde{v}}\epsilon,\quad \frac{\kappa^2 c^4}{a^2}\Xi^2\, .\ee
We distinguish between the coefficients $\kappa\overline\rho c^2/a$ and $\mathcal{H}^2$: they evolve synchronously during the matter-dominated stage of the
Universe evolution, but asynchronously during the $\Lambda$-dominated stage. The essence of the perturbative computation lies in the fact that for each term of
the second type in \rf{68b} there must exist a counterpart of the first type in \rf{67b}, such that their ratio is of the order of smallness $\epsilon$. This is
what we intend to confirm right now.

It can be easily seen that first five terms in \rf{68b}, divided by the corresponding first five terms in \rf{67b}, give precisely the order of smallness
$\epsilon$. Obviously, the same applies to the sixth terms. Indeed, $\rho\tilde{v}^2\ll|\delta\rho|$ at arbitrary distances \citep{Chisari}, and the helpful
estimate $\tilde v^2\sim\Phi\delta\rho/\rho$ (see \cite{Baumann}) holds true. Hence, $\rho\tilde v^2/\delta\rho\sim\Phi\sim\epsilon$. Further, since $\kappa
c^2\Xi/a \sim \Phi'+\mathcal{H}\Phi$ \rf{11a} and $\kappa c^2\rho{\bf\tilde v}/a \sim \triangle{\bf B}-2\kappa\overline{\rho}c^2{\bf B}/a$ \rf{9a}, the seventh
term in \rf{68b} is reduced to a combination of $(\mathcal{H}/\Upsilon)\epsilon^2$ and $\mathcal{H}^2\epsilon^2$ while the eighth term is reduced to a combination
of $\epsilon^2/L^2$ and $\left(\kappa\overline{\rho}c^2/a\right)\epsilon^2$. This quartet is already present in \rf{68b}, hence, the seventh and eighth summands
add nothing new. Similarly, the last term $\kappa^2c^4\Xi^2/a^2 \sim (\Phi'+\mathcal{H}\Phi)^2 = \Phi'^2+2\mathcal{H}\Phi\Phi'+\mathcal{H}^2\Phi^2$. This is a
combination of $\epsilon^2/\Upsilon^2$, $(\mathcal{H}/\Upsilon)\epsilon^2$ and $\mathcal{H}^2\epsilon^2$. Further,
$\Phi''=-3\mathcal{H}\Phi'-\left(2\mathcal{H}'+\mathcal{H}^2\right)\Phi$ \rf{11a} and $\mathcal{H}'=\mathcal{H}^2-\kappa\overline\rho c^2/(2a)$ \rf{2a}, hence, in
its turn, $\epsilon^2/\Upsilon^2$ may be treated as a combination of $(\mathcal{H}/\Upsilon)\epsilon^2$, $\mathcal{H}^2\epsilon^2$ and
$\left(\kappa\overline{\rho}c^2/a\right)\epsilon^2$. Consequently, the last summand in \rf{68b} also adds nothing new to those terms which are already available
in the collection.

Thus, we have shown that the elaborated perturbative scheme is valid. This scheme elegantly resolves the formidable challenge briefly discussed in the
introductory part of \citep{Clarkson1}: at any cosmological scale for each summand in the equations for the second-order metric corrections there exists a much
larger counterpart in the corresponding equations for the first-order metric corrections. Therefore, in particular, the situation when magnitudes of $\Phi^{(2)}$
and $\Phi$ are comparable is really improbable. Quite the contrary, the inequality $\left|\Phi^{(2)}\right|\ll|\Phi|$ may be expected to occur everywhere, as it
certainly should be in the framework of a self-consistent perturbation theory.

\

\subsection{Minkowski Background Limit}\label{subsec35}

In this subsection, again for the sake of simplicity, we momentarily ignore all terms being quadratic in particle velocities and concentrate on the Minkowski
background limit: the scale factor $a$ is now just a constant, $\mathcal H=0$, $\overline\rho=0$. Then, according to \cite{Eingorn},
\be{69b} \Phi=-\frac{\kappa c^2}{8\pi a}\sum\limits_n\frac{m_n}{|{\bf r}-{\bf r}_n|}\, ,\ee
\be{70b} {\bf B}=\frac{\kappa c^2}{4\pi a}\sum\limits_{n}\left[\frac{m_n{\bf\tilde v}_n}{|{\bf r}-{\bf r}_n|}+\frac{m_n[{\bf\tilde v}_n({\bf r}-{\bf r}_n)]}{|{\bf
r}-{\bf r}_n|^3}({\bf r}-{\bf r}_n)\right]\, .\ee
The sum of Newtonian potentials \rf{69b} is a solution of the standard Poisson equation
\be{71b} \triangle\Phi=\frac{\kappa c^2}{2a}\rho=\frac{\kappa c^2}{2a}\sum\limits_{n}m_n \delta({\bf r}-{\bf r}_n)\, . \ee

At the same time, from the second equation in \rf{56b} and \rf{30b} we get
\be{72b} \triangle\Psi^{(2)}=-\Phi\triangle\Phi-\frac{3}{2}\left(\nabla\Phi\right)^2=-\frac{3}{4}\triangle\left(\Phi^2\right)+\frac{\kappa c^2}{4a}\rho\Phi\, ,\ee
where an evident relationship $2\left(\nabla\Phi\right)^2=\triangle\left(\Phi^2\right)-2\Phi\triangle\Phi$ has been used along with \rf{71b}. Hence,
\be{73b} \Psi^{(2)}=-\frac{3}{4}\Phi^2 - \frac{\kappa c^2}{16\pi a}\sum\limits_{n}\frac{m_n}{|{\bf r}-{\bf r}_n|}\left.\Phi\right|_{{\bf r}={\bf r}_n}\, .\ee
After lengthy calculation, being based on \rf{32b}, \rf{33b}, \rf{35b} and \rf{42b}, one also finds
\be{74b} Q^{(\mathrm{S})} = \frac{7}{4a^2}\Phi^2 - \frac{\kappa c^2}{16\pi a^3}\sum\limits_{n}\frac{m_n}{|{\bf r}-{\bf r}_n|}\left.\Phi\right|_{{\bf r}={\bf r}_n}
+\frac{3\kappa c^2}{16\pi a^3}\sum\limits_{n}m_n\frac{({\bf r}-{\bf r}_n)}{|{\bf r}-{\bf r}_n|}\left.(\nabla\Phi)\right|_{{\bf r}={\bf r}_n}\, . \ee
Substitution of \rf{73b} and \rf{74b} into the first equation in \rf{56b} gives
\be{75b} \Phi^{(2)} = \Phi^2 - \frac{\kappa c^2}{8\pi a}\sum\limits_{n}\frac{m_n}{|{\bf r}-{\bf r}_n|}\left.\Phi\right|_{{\bf r}={\bf r}_n} +\frac{3\kappa
c^2}{16\pi a}\sum\limits_{n}m_n\frac{({\bf r}-{\bf r}_n)}{|{\bf r}-{\bf r}_n|}\left.(\nabla\Phi)\right|_{{\bf r}={\bf r}_n}\, .\ee
As usual, the gravitational field produced by the $n$-th particle is excluded from the factors $\left.\Phi\right|_{{\bf r}={\bf r}_n}$ and
$\left.(\nabla\Phi)\right|_{{\bf r}={\bf r}_n}$.

Let us compare the solutions \rf{70b} and \rf{75b} with the corresponding adapted expressions
\be{76b} {\bf B}_{\mathrm{LL}}=\frac{\kappa c^2}{16\pi a}\sum\limits_{n}\left[7\frac{m_n{\bf\tilde v}_n}{|{\bf r}-{\bf r}_n|}+\frac{m_n[{\bf\tilde v}_n({\bf
r}-{\bf r}_n)]}{|{\bf r}-{\bf r}_n|^3}({\bf r}-{\bf r}_n)\right]\, ,\ee
%
\be{77b} \Phi^{(2)}_{\mathrm{LL}} = \Phi^2 - \frac{\kappa c^2}{8\pi a}\sum\limits_{n}\frac{m_n}{|{\bf r}-{\bf r}_n|}\left.\Phi\right|_{{\bf r}={\bf r}_n}\, ,\ee
which are equivalent to those from the textbook by \cite{Landau} (see the formulas (106.15) and (106.13) therein). Here we still ignore velocities squared as we
arranged before. Of course, neither \rf{70b} coincides with \rf{76b}, nor \rf{75b} coincides with \rf{77b}. As pointed out by \cite{Eingorn}, the reason lies in
the fact that our gauge conditions differ from those applied by \cite{Landau}. Therefore, in order to reach agreement with this textbook, suffice it to find such
a transformation of coordinates that would establish desired linkage. Apparently, it is enough to transform only the temporal coordinate:
$\eta\mapsto\eta-A(\eta,{\bf r})$, then $\Phi^{(2)}\mapsto\Phi^{(2)}+A'$ and ${\bf B}\mapsto{\bf B}+\nabla A$. Demanding that
\be{78b} \Phi^{(2)}+A'=\Phi^{(2)}_{\mathrm{LL}},\quad {\bf B}+\nabla A={\bf B}_{\mathrm{LL}}\, ,\ee
with the help of \rf{70b}, \rf{75b}, \rf{76b} and \rf{77b} we get
\be{79b} A'=-\frac{3\kappa c^2}{16\pi a}\sum\limits_{n}m_n\frac{({\bf r}-{\bf r}_n)}{|{\bf r}-{\bf r}_n|}\left.(\nabla\Phi)\right|_{{\bf r}={\bf r}_n}\, ,\ee
\be{80b} \nabla A = \frac{3\kappa c^2}{16\pi a}\sum\limits_{n}\left[\frac{m_n{\bf\tilde v}_n}{|{\bf r}-{\bf r}_n|}-\frac{m_n[{\bf\tilde v}_n({\bf r}-{\bf
r}_n)]}{|{\bf r}-{\bf r}_n|^3}({\bf r}-{\bf r}_n)\right]\, .\ee
Action of $\nabla$ on both sides of \rf{79b} gives
\ba{81b} \nabla A'=-\frac{3\kappa c^2}{16\pi a}\sum\limits_{n}\frac{m_n}{|{\bf r}-{\bf r}_n|}\left[\left.(\nabla\Phi)\right|_{{\bf r}={\bf r}_n}-\frac{\left[({\bf
r}-{\bf r}_n)\left.(\nabla\Phi)\right|_{{\bf r}={\bf r}_n}\right]}{|{\bf r}-{\bf r}_n|}\frac{({\bf r}-{\bf r}_n)}{|{\bf r}-{\bf r}_n|}\right]\, ,\ea
and exactly the same result follows also from \rf{80b}. This incontestable fact ensures existence of the function $A(\eta,{\bf r})$ and, consequently, of the
above-mentioned coordinate transformation. Thus, agreement with \cite{Landau} has been reached.

\

\section{AVERAGING INITIATIVES ON THE EVE OF COSMOLOGICAL BACKREACTION ESTIMATION}\label{sec4}

\setcounter{equation}{0}

In view of the predictably zero average values of the first-order metric corrections \citep{Eingorn}, the computation of the cosmological backreaction effects
should be based on the second-order perturbation theory. Without pretending to an exhaustive study, let us perform the Euclidean averaging, or smoothing
\citep{Clarkson2}, of the $00$-component of Einstein equations \rf{36b}, multiplied by $a^2/2$, and the sum of $11$-, $22$- and $33$-components (see \rf{38b}),
multiplied by $\left(-a^2/6\right)$. We gather all terms containing $\overline{\Psi^{(2)}}$, $\overline{\Phi^{(2)}}$ and their temporal derivatives in the lhs,
while the other averaged contributions are gathered in the rhs:
\be{1c} -3\mathcal{H}\overline{\Psi^{(2)}}'-3\mathcal{H}^2\overline{\Phi^{(2)}}-\frac{3\kappa\overline\rho
c^2}{2a}\overline{\Psi^{(2)}}=\frac{1}{2}a^2\overline{Q_{00}} \equiv \frac{1}{2}\kappa a^2\overline{\varepsilon}^{(\mathrm{II})}\, ,\ee
\be{2c} \overline{\Psi^{(2)}}''+\mathcal{H}\left(2\overline{\Psi^{(2)}}+\overline{\Phi^{(2)}}\right)'
+\left(2\mathcal{H}'+\mathcal{H}^2\right)\overline{\Phi^{(2)}}=-\frac{1}{6}a^2\overline{Q_{\alpha\alpha}} \equiv \frac{1}{2}\kappa
a^2\overline{p}^{(\mathrm{II})}\, .\ee
Here the overline indicates integrating over a comoving volume $\mathcal{V}$ and dividing by this volume in the limit of the infinite integration domain
($\mathcal{V}\rightarrow+\infty$). In addition, we have introduced the effective average energy density $\overline{\varepsilon}^{(\mathrm{II})}(\eta)$ and
pressure $\overline{p}^{(\mathrm{II})}(\eta)$:
\ba{3c} \kappa\overline{\varepsilon}^{(\mathrm{II})}&\equiv& \overline{Q_{00}}= \frac{\kappa c^2}{2a^3}\overline{\rho\tilde{v}^2}- \frac{3\kappa^2
c^4}{4a^4}\overline{\Xi^2}+\frac{6\kappa c^2}{a^3}\mathcal{H}\overline{\Xi\Phi} -\left(\frac{3\kappa\overline{\rho}c^2}{2a^3}
+\frac{15}{a^2}\mathcal{H}^2\right)\overline{\Phi^2}\nn\\
&+&\left(\frac{\kappa\overline{\rho} c^2}{2a^3}+\frac{3}{a^2}\mathcal{H}^2\right)\overline{{\bf
B}^2}-\frac{2}{a^2}\overline{\Phi\triangle\Phi}-\frac{3}{a^2}\overline{(\nabla\Phi)^2}-\frac{1}{4a^2}\overline{{\bf B}\triangle{\bf B}} -\frac{\kappa
c^2}{a^3}\overline{\rho{\bf \tilde v}{\bf B}}\, ,\ea
\ba{4c} -3\kappa\overline{p}^{(\mathrm{II})} &\equiv& \overline{Q_{\alpha\alpha}}=-\frac{\kappa c^2}{a^3}\overline{\rho \tilde{v}^2}-\frac{3\kappa^2
c^4}{4a^4}\overline{\Xi^2}+\frac{9\kappa c^2}{a^3}\mathcal{H}\overline{\Xi\Phi}+\left(\frac{12\kappa\overline\rho
c^2}{a^3}-\frac{15}{a^2}\mathcal{H}^2\right)\overline{\Phi^2}\nn \\
&-&\frac{3}{a^2}\mathcal{H}^2\overline{{\bf B}^2} -\frac{8}{a^2}\overline{\Phi\triangle\Phi} -\frac{7}{a^2}\overline{(\nabla\Phi)^2} -\frac{5}{4a^2}\overline{{\bf
B}\triangle{\bf B}} -\frac{2\kappa c^2}{a^3}\overline{\rho{\bf \tilde v}{\bf B}}\, ,\ea
where the explicit expressions \rf{30b} for $Q_{00}$ and \rf{35b} for $Q_{\alpha\alpha}$ have been used. Replacing $\overline{(\nabla\Phi)^2}$ by
$-\overline{\Phi\triangle\Phi}$ and expressing $\triangle\Phi$ and $\triangle{\bf B}$ from \rf{8a} and \rf{9a}, respectively, we rewrite \rf{3c} and \rf{4c} in
the more compact form:
\be{5c} \kappa\overline{\varepsilon}^{(\mathrm{II})} = \frac{\kappa c^2}{2a^3}\overline{\rho\tilde{v}^2}+\frac{\kappa c^2}{2a^3}\overline{\rho\Phi}-
\frac{3\kappa^2 c^4}{4a^4}\overline{\Xi^2}+\frac{9\kappa c^2}{2a^3}\mathcal{H}\overline{\Xi\Phi}
-\frac{15}{a^2}\mathcal{H}^2\overline{\Phi^2}+\frac{3}{a^2}\mathcal{H}^2\overline{{\bf B}^2} -\frac{\kappa c^2}{2a^3}\overline{\rho{\bf \tilde v}{\bf B}} \, ,\ee
\ba{6c} \kappa\overline{p}^{(\mathrm{II})} &=& \frac{\kappa c^2}{3a^3}\overline{\rho \tilde{v}^2}+\frac{\kappa c^2}{6a^3}\overline{\rho\Phi}+\frac{\kappa^2
c^4}{4a^4}\overline{\Xi^2}-\frac{7\kappa c^2}{2a^3}\mathcal{H}\overline{\Xi\Phi}-\left(\frac{7\kappa\overline{\rho}
c^2}{2a^3}-\frac{5}{a^2}\mathcal{H}^2\right)\overline{\Phi^2}\nn\\
&+&\left(\frac{5\kappa\overline{\rho} c^2}{6a^3}+\frac{1}{a^2}\mathcal{H}^2\right)\overline{{\bf B}^2}-\frac{\kappa c^2}{6a^3}\overline{\rho{\bf \tilde v}{\bf B}}
\, .\ea

Expressing $\overline{\varepsilon}^{(\mathrm{II})}$ and $\overline{p}^{(\mathrm{II})}$ from \rf{1c} and \rf{2c}, one can easily verify that these functions
satisfy the standard conservation equation
\be{7c} \left(a^3\overline{\varepsilon}^{(\mathrm{II})}\right)' +3a^3\mathcal{H}\overline{p}^{(\mathrm{II})}=0\, ,\ee
as it certainly should be. Hence, the expressions \rf{5c} and \rf{6c} for the same functions must automatically satisfy this equation as well. This can be
verified through the instrumentality of the formulas \rf{59b} and \rf{60b} as well as equations from Section~\ref{sec2}.

It is worth mentioning that if one keeps in the rhs of \rf{5c} and \rf{6c} only first two terms, which dominate at sufficiently small scales, and makes use of the
relationship $\overline{\rho\Phi}=-2\overline{\rho\tilde{v}^2}$, which holds true for the virialized regions, then $\overline{p}^{(\mathrm{II})}\rightarrow0$
while $\overline{\varepsilon}^{(\mathrm{II})}\rightarrow-\left[c^2/\left(2a^3\right)\right]\overline{\rho\tilde{v}^2}\sim 1/a^3$. Thus, at virialized scales the
effective pressure $\overline{p}^{(\mathrm{II})}$ vanishes while the non-vanishing effective average energy density $\overline{\varepsilon}^{(\mathrm{II})}$
brings to a small time-independent renormalization of the corresponding background quantity $\overline\varepsilon$, in full accord with \cite{Baumann} (see
additionally \cite{Wetterich} for earlier theoretical efforts and a ``cosmic virial theorem''). The interpretation of the simulation outputs by \cite{Adamek} also
suggests that ``stable clustering'' (implying virialized nonlinear structures) razes backreaction from the cosmological battlefield. The underlying perturbative
scheme advocated by \cite{Adamek} is compared with ours by \cite{Eingorn}. It is necessary to mention that this purely numerical scheme is characterized by the
first order accuracy for large enough distances and the second order accuracy for sufficiently small distances, while the approach advocated in the current paper
is characterized by the second order accuracy everywhere and is fully analytical at least with respect to the first-order cosmological perturbations \rf{5a};
\rf{6a} and the sources \rf{30b}--\rf{33b} of the second-order ones.

It is noteworthy as well that the velocity-dependent summands can be easily distinguished from the velocity-independent ones in \rf{5c} and \rf{6c}, and there are
only two types of contributions, which do not contain particle velocities: $\sim\overline{\rho\Phi_0}$ and $\sim\overline{\Phi_0^2}$. Here $\Phi_0$ denotes the
velocity-independent part of the first-order scalar perturbation $\Phi$ \rf{5a}, that is the sum of Yukawa potentials with the same interaction range $\lambda$
(up to an additive constant $1/3$):
\be{8c} \Phi_0=\frac{1}{3}-\frac{\kappa c^2}{8\pi a} \sum\limits_{n}\frac{m_n}{|{\bf r}-{\bf r}_n|}\exp\left(-\frac{a|{\bf r}-{\bf r}_n|}{\lambda}\right)\, .\ee
For illustration purposes, we compute both average quantities $\overline{\rho\Phi_0}$ and $\overline{\Phi_0^2}$ analytically:
\be{9c} \overline{\rho\Phi_0} = \frac{1}{3}\overline{\rho}-\frac{\kappa c^2}{8\pi a}\,\frac{1}{\mathcal{V}}\sum\limits_{n}\sum\limits_{k\neq n}\frac{m_n
m_k}{|{\bf r}_k-{\bf r}_n|}\exp\left(-\frac{a|{\bf r}_k-{\bf r}_n|}{\lambda}\right)\, ,\ee
\be{10c} \overline{\Phi_0^2} = -\frac{1}{9} +\frac{\kappa c^2}{48\pi\overline{\rho}\lambda}\,\frac{1}{\mathcal{V}} \sum\limits_{n}\sum\limits_{k}m_n m_k
\exp\left(-\frac{a|{\bf r}_k-{\bf r}_n|}{\lambda}\right)\, .\ee
It presents no difficulty to receive evidence that both these expressions tend to zero in the homogeneous mass distribution limit ($\sum\sum
m_nm_k\rightarrow\overline{\rho}^2\iint d{\bf r}_nd{\bf r}_k$), as it certainly should be since $\Phi_0=0$ at any point in this test limit \citep{Eingorn}.

Reverting to \rf{5c} and \rf{6c}, we emphasize that collections of terms in the right-hand sides may assist in the cosmological backreaction estimation. We
formulate the following quite feasible two-stage plan:

\vspace{0.2cm}

\ $\bullet$ \ the launch of a new generation of cosmological $N$-body simulations based on the formalism developed by \cite{Eingorn} (see the equations of motion
(3.6) therein);

\vspace{0.2cm}

\ $\bullet$ \ the use of outputs of these simulations for the estimation of the effective average energy density $\overline{\varepsilon}^{(\mathrm{II})}$ and
pressure $\overline{p}^{(\mathrm{II})}$, and the subsequent comparison with the background quantity $\overline\varepsilon$.

\vspace{0.2cm}

If the underlying inequalities $\left|\overline{\varepsilon}^{(\mathrm{II})}\right|\ll \overline\varepsilon$ and $\left|\overline{p}^{(\mathrm{II})}\right|\ll
\overline\varepsilon$ become doubtful at any moment during the matter-dominated or $\Lambda$-dominated stages of the Universe evolution, then this fact may serve
as a sure sign of backreaction significance and inappropriateness of the FLRW metric \rf{1a} and Friedmann equations \rf{2a}, with inevitable grave consequences.
At the same time, if the inequalities being tested seem always unquestionable, this result by itself does not necessarily mean that backreaction is insignificant,
for the simple reason that we still rely on the initial assumption of the FLRW background existence and actually have no predictive power beyond. Nevertheless, if
this key assumption is valid and $\overline\varepsilon$ is really much greater than $\left|\overline{\varepsilon}^{(\mathrm{II})}\right|$ and
$\left|\overline{p}^{(\mathrm{II})}\right|$, then we can formally celebrate the preliminary success of the elaborated perturbative approach and add a third stage
relying on \rf{55b} and \rf{56b}:

\vspace{0.2cm}

\ $\bullet$ \ the estimation of $\Phi^{(2)}$, $\Psi^{(2)}$ and ${\bf B}^{(2)}$, and the subsequent comparison with $\Phi$ and ${\bf B}$, respectively.

\vspace{0.2cm}

The proposed plan exploits Yukawa gravity resulting from GR \citep{Eingorn,Eingorn2017} and therefore possesses a definite advantage over generally accepted
simulations exploiting Newtonian gravity \citep{Peebles,Springel}. This advantage is clearly explained by \cite{Rasanen2010,Rasanen}: in the framework of the
so-called Newtonian cosmology, as opposed to GR, the backreaction effects are reduced to boundary terms vanishing in the case of the standard periodic boundary
conditions. It is stressed by \cite{Rasanen2010,Rasanen} that Newtonian gravity does not represent the weak field limit of GR. We emphasize that the formulation
of this crucial limit is no longer an open issue: the corresponding cosmological perturbation theory incorporating nonlinear density contrasts has been developed
by \cite{Eingorn} and extended in the current paper. As pointed out by \cite{Eingorn}, Yukawa screening of the interparticle attraction may be treated as a
relativistic effect. This is especially important in view of the fact that backreaction significance is inseparably linked with non-Newtonian gravitational
physics \citep{Rasanen2010,Buchert1}.


Of course, the first-order scheme elaborated by \cite{Eingorn} is insufficient for $N$-body simulations taking into consideration the second-order metric
corrections in the equations of motion, and the formalism developed in the current paper should be employed instead. This is an indispensable complication of the
aforesaid simplified plan from the point of view of modern precision cosmology. There are formidable obstacles on the way of solving the derived equations
\rf{55b} and \rf{56b} analytically, however, owing to their linearity with respect to the sought-for functions, it presents no difficulty in principle to solve
these equations numerically. Furthermore, it would be very interesting to study propagation of gravitational waves in the inhomogeneous Universe, governed by
\rf{47b}. Having at our disposal the determined first- and second-order cosmological perturbations, we can in principle introduce and calculate the ``averaged
Hubble rate'' as well as the kinematical and dynamical backreactions \citep{Behrend,Brown2,Brown1}. Moreover, the thorough analysis of the ``effective equation of
state'' (that is the relationship between $\overline\varepsilon$, $\overline{\varepsilon}^{(\mathrm{II})}$ and $\overline{p}^{(\mathrm{II})}$) for different
cosmological epochs becomes feasible (see additionally \cite{Li1,Li2,Schwarz,Bose} for related theoretical efforts). It is also noteworthy that all formulas,
where the discreteness of matter is not manifestly specified (for example, by means of the delta-functions), are valid for continuous medium as well. Thus, the
transition to the hydrodynamical description is simple.

Implementation of both aforesaid ``simplified'' and ``complicated'' plans is extremely promising, but oversteps the limits of our prolonged narration.

\

\section{CONCLUSION}\label{sec5}

Summarizing, we recollect the main results achieved in the current paper within the conventional $\Lambda$CDM model:

\vspace{0.2cm}

\ $\bullet$ \ the equations \rf{56b}, \rf{55b} and \rf{47b} for the second-order scalar ($\Phi^{(2)}$, $\Psi^{(2)}$), vector (${\bf B}^{(2)}$) and tensor
($h_{\alpha\beta}$) cosmological perturbations have been derived. These equations are suitable at all spatial scales (naturally, except for the regions of strong
gravitational fields in immediate proximity to such their generators as black holes or neutron stars) and permit of nonlinear density contrasts;

\vspace{0.2cm}

\ $\bullet$ \ the Helmholtz equations \rf{55b} and \rf{56b} incorporate exactly the same Yukawa interaction ranges as their corresponding counterparts \rf{9a} and
\rf{8a} determining the first-order metric corrections;

\vspace{0.2cm}

\ $\bullet$ \ the constructed scheme passes three important corroborative tests:

{\parindent=14mm 1) we have verified that all Einstein equations are satisfied within the adopted accuracy along with the gauge conditions
(subsection~\ref{subsec33});}

{\parindent=14mm 2) we have confirmed the self-consistency of order assignments and the related expectation that the first-order metric corrections dominate over
the second-order ones everywhere (subsection~\ref{subsec34});}

{\parindent=14mm 3) in the Minkowski background limit the linkage with the textbook material has been established (subsection~\ref{subsec35});}

\vspace{0.2cm}

\ $\bullet$ \ the highway to investigate the cosmological backreaction effects beyond Newtonian gravitational physics has been outlined. The proposed relativistic
simulations of the cosmic structure growth accompanied by the investigations of propagation of light and gravitational waves in the inhomogeneous Universe would
definitely assist in deepening and testing our knowledge of the spacetime and world's filling material including dark ingredients.

\vspace{0.1cm}

\section*{ACKNOWLEDGEMENTS}

We are grateful to the anonymous Referee for the valuable comments. The work of R.~Brilenkov was partially supported by the EMJMD Student Scholarship from the
Erasmus\,+\,: Erasmus Mundus Joint Master Degree programme AstroMundus in Astrophysics. The work of M.~Eingorn was partially supported by NSF CREST award
HRD-1345219 and NASA grant NNX09AV07A.


\vspace{0.1cm}

\begin{thebibliography}{}

%
\bibitem[Adamek et~al.(2015)]{Adamek}
Adamek, J., Clarkson, C., Durrer, R., \& Kunz, M.
2015, PhRvL, 114, 051302
%
%
\bibitem[Ade et~al.(2016)]{Planck15}
Ade, P. A. R., Aghanim, N., Arnaud, M., et al. [Planck Collaboration]
2016, A\&A, 594, A13
%
%
\bibitem[Amendola et~al.(2016)]{Euclid2}
Amendola, L., Appleby, S., Avgoustidis, A., et al.
2016, arXiv:1606.00180 [astro-ph.CO]
%
%
\bibitem[Bardeen(1980)]{Bardeen}
Bardeen, J. M.
1980, PhRvD, 22, 1882
%
%
\bibitem[Bartolo et~al.(2006)]{Bartolo}
Bartolo, N., Matarrese, S., \& Riotto, A.
2006, JCAP, 05, 010
%
%
\bibitem[Baumann et~al.(2012)]{Baumann}
Baumann, D., Nicolis, A., Senatore, L., \& Zaldarriaga, M.
2012, JCAP, 07, 051
%
%
\bibitem[Behrend et~al.(2008)]{Behrend}
Behrend, J., Brown, I. A., \& Robbers, G.
2008, JCAP, 01, 013
%
%
\bibitem[Ben-Dayan et~al.(2013)]{observ1}
Ben-Dayan, I., Gasperini, M., Marozzi, G., Nugier, F., \& Veneziano, G.
2013, JCAP, 06, 002
%
%
\bibitem[Bolejko \& Korzy\'{n}ski(2017)]{Bolejko}
Bolejko, K., \& Korzy\'{n}ski, M.
2017, IJMPD, 26, 1730011
%
%
\bibitem[Bonvin et~al.(2015a)]{observ2}
Bonvin, C., Clarkson, C., Durrer, R., Maartens, R., \& Umeh, O.
2015, JCAP, 06, 050
%
%
\bibitem[Bonvin et~al.(2015b)]{observ3}
Bonvin, C., Clarkson, C., Durrer, R., Maartens, R., \& Umeh, O.
2015, JCAP, 07, 040
%



%
\bibitem[Bose \& Majumdar(2013)]{Bose}
   Bose, N., \& Majumdar, A. S.
2013, GReGr, 45, 1971
%



%
\bibitem[Brown et~al.(2009)]{Brown2}
Brown, I. A., Behrend, J., \& Malik, K. A.
2009, JCAP, 11, 027
%
%
\bibitem[Brown et~al.(2013)]{Brown1}
Brown, I. A., Latta, J., \& Coley, A.
2013, PhRvD, 87, 043518
%
%
\bibitem[Buchert et~al.(2015)]{Buchert2}
Buchert, T., Carfora, M., Ellis, G. F. R., et al.
2015, CQGra, 32, 215021
%
%
\bibitem[Buchert \& R\"{a}s\"{a}nen(2012)]{Buchert1}
Buchert, T., \& R\"{a}s\"{a}nen, S.
2012, ARNPS, 62, 57
%
%
\bibitem[Chisari \& Zaldarriaga(2011)]{Chisari}
Chisari, N. E., \& Zaldarriaga, M.
2011, PhRvD, 83, 123505
%
%
\bibitem[Clarkson et~al.(2011)]{Clarkson2}
Clarkson, C., Ellis, G., Larena, J., \& Umeh, O.
2011, RPPh, 74, 112901
%
%
\bibitem[Clarkson \& Umeh(2011)]{Clarkson1}
Clarkson, C., \& Umeh, O.
2011, CQGra, 28, 164010
%
%
\bibitem[Durrer(2008)]{Ruth}
Durrer, R. 2008, The Cosmic Microwave Background (Cambridge: Cambridge University Press)
%
%
\bibitem[Eingorn(2016)]{Eingorn}
Eingorn, M.
2016, ApJ, 825, 84
%
%
\bibitem[Eingorn(2017)]{Eingorn2017}
Eingorn, M.
2017, IJMPD, 26, 1750121
%
%
\bibitem[Eingorn \& Brilenkov(2015)]{Ruslan}
Eingorn, M., \& Brilenkov, R.
2017, PDU, 17, 63
%
%
\bibitem[Eingorn et~al.(2016)]{ekz}
Eingorn, M., Kiefer, C., \& Zhuk, A.
2016, JCAP, 09, 032
%
%
\bibitem[Eingorn \& Zhuk(2014)]{EZcosm2}
Eingorn, M., \& Zhuk, A.
2014, JCAP, 05, 024
%
%
\bibitem[Gorbunov \& Rubakov(2011)]{Rubakov}
Gorbunov, D. S., \& Rubakov, V. A. 2011, Introduction to the Theory of the Early Universe: Cosmological Perturbations and Inflationary Theory (Singapore: World
Scientific)
%
%
\bibitem[Green \& Wald(2013)]{Green}
Green, S. R., \& Wald, R. M.
2013, PhRvD, 87, 124037
%
%
\bibitem[Landau \& Lifshitz(2000)]{Landau}
Landau, L. D., \& Lifshitz, E. M. 2000, The Classical Theory of Fields 
(Oxford: Oxford Pergamon Press)
%



%
\bibitem[Li \& Schwarz(2007)]{Li1}
   Li, N., \& Schwarz, D. J.
2007, PhRvD, 76, 083011
%
%
\bibitem[Li \& Schwarz(2008)]{Li2}
   Li, N., \& Schwarz, D. J.
2008, PhRvD, 78, 083531
%
%
\bibitem[Mukhanov(2005)]{Mukhanov}
   Mukhanov, V. 2005, Physical Foundations of Cosmology (Cambridge: Cambridge University Press)
%



%
\bibitem[Peebles(1980)]{Peebles}
Peebles, P. J. E. 1980, The large-scale structure of the Universe (Princeton: Princeton University Press)
%
%
\bibitem[R\"{a}s\"{a}nen(2010)]{Rasanen2010}
R\"{a}s\"{a}nen, S.
2010, PhRvD, 81, 103512
%
%
\bibitem[R\"{a}s\"{a}nen(2011)]{Rasanen}
R\"{a}s\"{a}nen, S.
2011, CQGra, 28, 164008
%
%
\bibitem[Scaramella et~al.(2015)]{Euclid1}
Scaramella, R., Mellier, Y., Amiaux, J., et al. [Euclid Collaboration]
2015, Proceedings of the International Astronomical Union, 10, 375
%



%
\bibitem[Schwarz(2010)]{Schwarz}
   Schwarz, D. J.
2010, arXiv:1003.3026 [astro-ph.CO]
%



%
\bibitem[Springel(2005)]{Springel}
Springel, V.
2005, MNRAS, 364, 1105
%
%
\bibitem[Wetterich(2003)]{Wetterich}
Wetterich, C.
2003, PhRvD, 67, 043513
%

\end{thebibliography}
\end{document}